\documentclass[twocolumn]{aastex63}

\newcommand{\ktwo}{{K2}}
\newcommand{\paperone}{{Paper I}}

\newcommand{\msun}{$M_{\odot}$}
\newcommand{\rsun}{$R_{\odot}$}

\received{July 20, 2021}
\revised{September 13, 2021}
\accepted{September 16, 2021}

\shorttitle{Pleiades, Praesepe and M35 Stellar Spins}
\shortauthors{Healy et al.}

\graphicspath{{./}}

\begin{document}

\title{Stellar Spins in the Pleiades, Praesepe, and M35 Open Clusters}

\correspondingauthor{Brian F. Healy}
\email{bfhealy@jhu.edu}

\author[0000-0002-7718-7884]{Brian F. Healy}
\affiliation{Department of Physics and Astronomy,\\
Johns Hopkins University \\
3400 North Charles Street, \\
Baltimore, MD 21218, USA}

\author[0000-0001-9165-9799]{P.R. McCullough}
\affiliation{Department of Physics and Astronomy,\\
Johns Hopkins University \\
3400 North Charles Street, \\
Baltimore, MD 21218, USA}
\affiliation{Space Telescope Science Institute, \\
3700 San Martin Drive, \\
Baltimore, MD 21218, USA}

\author[0000-0001-5761-6779]{Kevin C. Schlaufman}
\affiliation{Department of Physics and Astronomy,\\
Johns Hopkins University \\
3400 North Charles Street, \\
Baltimore, MD 21218, USA}


\begin{abstract}
We analyze spectroscopic and photometric data to determine the projected inclinations of stars in three open clusters: the Pleiades, Praesepe, and M35. We determine the $\sin i$ values of 42, 35, and 67 stars in each cluster, respectively, and from their distributions we find that isotropic spins and moderate alignment are both consistent with the Pleiades and Praesepe data. While it is difficult to distinguish between these scenarios for a single cluster, an ensemble of such distributions may facilitate a distinction. The M35 inclination distribution is most consistent with a superposition of isotropic and anisotropic spins, the source of which could be systematic error or a physical grouping of aligned stars. We also study internal cluster kinematics using radial velocities and proper motions. Our kinematics analysis reveals significant plane-of-sky rotation in Praesepe, with a mean velocity of $0.132 \pm 0.022$ km s$^{-1}$ in a clockwise direction. 
\end{abstract}


\section{Introduction} \label{sec:introduction}

Stars in open clusters offer unique insight into the outcomes of star formation through their assumed coeval formation. These stars' shared age, metallicity, and formation environment make each cluster a benchmark in temporal, chemical, and dynamical evolution. Among the quantities that offer a glimpse into a cluster's past is the projected inclination, the projected angle between a star's rotation axis and the observer's line of sight. An isotropic distribution of inclinations in a cluster
\citep[e.g.][]{jackson2010}
suggests a star-forming environment dominated by turbulent flow, while anisotropic groupings of spins
\citep[e.g.][]{corsaro2017}
present evidence for the inheritance and preservation of molecular cloud angular momentum by nascent cluster stars.

Magnetic fields may influence spin alignment in open clusters, as suggested by a study of proto-planetary disks that found a 2$\sigma$ deviation from isotropy among the position angles of disks in the molecular cloud Lupus III \citep[][]{masataka2020}. This work found a correlation between the disk orientations and magnetic field direction in this cloud. However, it also measured randomly oriented position angles in four other star-forming regions, suggesting a lesser influence of magnetic fields in those cases.

\citet[][hereafter \paperone]{paper1_healy} determined 33 inclinations in the open cluster NGC 2516, finding a distribution consistent with either isotropy or moderate alignment. Studying additional clusters in a uniform way will reveal whether this result is an outlier or representative of a larger sample. In order to learn from an ensemble of uniformly determined inclinations, in this paper we measure the spin-axis distributions of stars in the Pleiades, Praesepe, and M35 clusters (Table \ref{tab:cluster_basics}).

Two previous inclination studies of the Pleiades found that isotropic distributions were favored \citep[][]{jackson2010, jackson2018}. The Praesepe cluster has also been the subject of two studies, with one finding a preferred anisotropic distribution \citep{kovacs2018} and the other favoring isotropy \citep[][]{jackson2019}. There has not been a study of stellar inclinations in M35 to date.

We use the spectro-photometric method to determine projected inclinations, which is governed by the following equation:
\begin{equation}
    v\sin i = \frac{2\pi R}{P}\sin i,
    \label{eq:spectrophot}
\end{equation}
Here, $v\sin i$ is the projected rotation velocity measured from Doppler broadening of spectral lines, $R$ is the radius obtained through isochrone and spectral-energy-distribution (SED) fitting to stellar magnitudes, and $P$ the the rotation period observed as a modulating signal in time-series photometry. In this paper, we will often refer to the projected $\sin i$ value on the right-hand side of Eq. \ref{eq:spectrophot} as the ``inclination.'' Fitting models to the resulting inclination distributions reveals their tendency toward isotropy or alignment.

We also perform an analysis of proper motions and line-of-sight (LOS) velocities for each cluster in order to study their internal kinematics. This part of our study follows the approach and motivation of \citet[][]{kamann2019}, who explored possible connections between the aligned spins seen in two open clusters \citep[][]{corsaro2017} and these clusters' overall rotation.

The rest of the paper is structured as follows: Sec. \ref{sec:datasources} describes the data sources used in our analysis of the three clusters. Sec. \ref{sec:analysis} provides the details of our data analysis. Sec. \ref{sec:results} presents our results for each cluster. Sec. \ref{sec:discussion} discusses the implications of our results and we conclude in Sec. \ref{sec:conclusion}.

\begin{table*}[]
\begin{center}
\label{tab:cluster_basics}
    {\centering
    \begin{tabular}{c|c|c|c|c|c}
    \hline
    \hline
         Cluster & $l$ [$^{\circ}$] & $b$ [$^{\circ}$] & $d$ [pc] & age [Myr] & mass [\msun]  \\
         \hline
         Pleiades & 166.428 & -23.602 & $134.8 \pm 1.3$ & 110-160$^{1}$ & $\sim$ 800$^{4}$ \\
         Praesepe & 205.964 & 32.415 & $183.1 \pm 2.3$ & 600-800$^{2,1}$ & $\sim$ 630$^{5}$ \\
         M35 & 186.596 & 2.223 & $838 \pm 37 $ & $\sim$ 150$^{3}$ & $\sim$ 1600$^{6}$ \\
        \hline
    \end{tabular}
    }
\end{center}
    \footnotesize{$^{1}$ \citealp{gossage2018}, $^{2}$ \citealp{brandt2015}, $^{3}$ \citealp{meibom2009}, $^{4}$ \citealp{pleiades_mass}, $^{5}$ \citealp{praesepe_mass_function}, $^{6}$ \citealp{m35_mass_function}}
\begin{center}
\caption{Basic parameters for each cluster in this study, including galactic latitude, longitude, and distance from Gaia EDR3 \citep{gaia_edr3_2020} along with age and mass. References for age and mass are noted below the table.}
\end{center}
\end{table*}

\section{Data Sources} \label{sec:datasources}
For each cluster, we consulted the Gaia DR2-based \citep[][]{gaia2016, gaia2018} member lists from \citet{cantatgaudin2018}. For all subsequent analysis, we used the Gaia DR2-to-EDR3 match table to update target stars with EDR3 astrometry \citep[][]{gaia_edr3_2020}. Since Sun-like stars are most amenable to a spectro-photometric inclination determination, we did not use Gaia EDR3 data to identify additional cluster members. The majority of these newly identified members would likely be faint (low-mass) stars that would be removed by our analysis anyway. 

Using TOPCAT \citep[][]{taylor2005_topcat}, we performed 1'' cross-matches between the cluster member lists and all additional data sets that lacked pre-determined Gaia source IDs. These data included stellar magnitudes from CatWISE2020 \citep[][]{catwise_2020}, 2MASS \citep[][]{skrutskie2006}, Hipparcos/Tycho-2 \citep[][]{perryman1997, hoeg2000}, APASS DR10 \citep[][]{apass_dr10}, and GALEX \citep[][]{galex}. For the rotation period $P$ and $v\sin i$, we used data as noted in the following subsections.

\subsection{Pleiades and Praesepe}
For the Pleiades, we used rotation period measurements derived from \ktwo\ observations \citep[][]{k2} by \citet{rebull2016_pleiades_pap2}. These measurements are based on periodogram analyses using the NASA Exoplanet Archive Periodogram Service \citep[][]{akeson2013_periodogram_service}. We adopted conservative 1\% random uncertainties in each period, representing the least precise measurements \citetext{L. Rebull 2021, private communication}. We estimate the systematic effects of differential rotation in Sec. \ref{subsec:diffrot} We used $v\sin i$ measurements from the CORAVEL instrument, reported by \citet[][]{mermilliod2009}.

For Praesepe, we used \ktwo-based rotation period measurements from \citet{rebull2017_praesepe}, who employed the same analysis techniques as for the Pleiades periods. We once again adopted 1\% random uncertainties on each measurement. We used $v\sin i$ data from \citet{mermilliod2009}.

\subsection{M35}
For M35, we used \ktwo\ rotation period measurements from \citet{soares-furtado2020_m35rot} for stars with periodogram SNR $> 3$. We used the periodogram signal-to-noise ratio listed for each period to estimate uncertainties. We selected stars classified as rotating with no blending. 

The $v\sin i$ measurements for this cluster came from the WIYN Open Cluster Study \citep[WOCS;][]{geller2010, m35_vsini_leiner_2015}. To ensure that each $v\sin i$ measurement quantified the uncertainty in the parameter itself rather than its mean, we converted the given standard error uncertainties to standard deviations by multiplying by the square root of the number of observations. Finally, we accessed WOCS $BVRI$ magnitudes \citep[][]{m35_wocs_thompson2014} to supplement the other photometric surveys for this most distant cluster of our sample.

\section{Analysis} \label{sec:analysis}
Our analysis begins by removing stars that are most likely to yield spurious measurements, such as binary stars close to equal mass. The possibilities of $v\sin i$ being confused in double-lined spectra, multiple rotation signals to be present in a light curve, and excess brightness to be seen in an SED justify removing them. After establishing a set made up of stars presumed to be either single or binaries dominated by the primary, we perform isochrone/SED fitting to determine radii and combine these results with published period and $v\sin i$ measurements with a Bayesian method to determine $\sin i$. We model the resulting cumulative distribution functions (CDFs) of $\sin i$ values using Monte Carlo methods and we also analyze proper motion and LOS velocity measurements to learn about each cluster's internal kinematics. 

\subsection{Star selection}
\label{subsec:member_selection}

We initially selected stars having a minimum membership probability threshold of 68\%. Following the approach of \paperone, we removed likely equal-mass binaries from each cluster's color-magnitude diagram along with systems classified as binaries by SIMBAD \citep[][]{wenger2000}. We excluded four additional stars from the M35 sample that were indicated to be binaries by \citet[][]{m35_vsini_leiner_2015}.

To further reduce the number of likely binaries in our sample, we analyzed the Gaia EDR3 Reduced Unit Weight Error (RUWE) measurement using a technique motivated by \citet[][]{belokurov2020}. We first performed a Gaussian kernel density estimation on the distribution of RUWE values (with a bandwidth of 0.2) for each cluster to create a probability density function (PDF). We isolated the portion of the PDF corresponding to values $\leq$ the RUWE value at the peak of the PDF, which for each cluster is $\sim 1$. We then reflected this curve about the maximum and normalized it, creating a RUWE PDF representative of likely single stars. We discarded the stars with RUWE values that were greater than the 99th percentile of this single-star distribution.

\subsection{Isochrone/SED fitting}
To determine stellar radii, effective temperatures, and other parameters, we used the \texttt{isochrones} software \citep[][]{morton2015} to fit MIST isochrones \citep[][]{choi2016} and their predicted flux values to each star's broad-band SED and Gaia EDR3 parallax measurement.

\begin{figure*}
    \centering
    \includegraphics[scale=0.45]{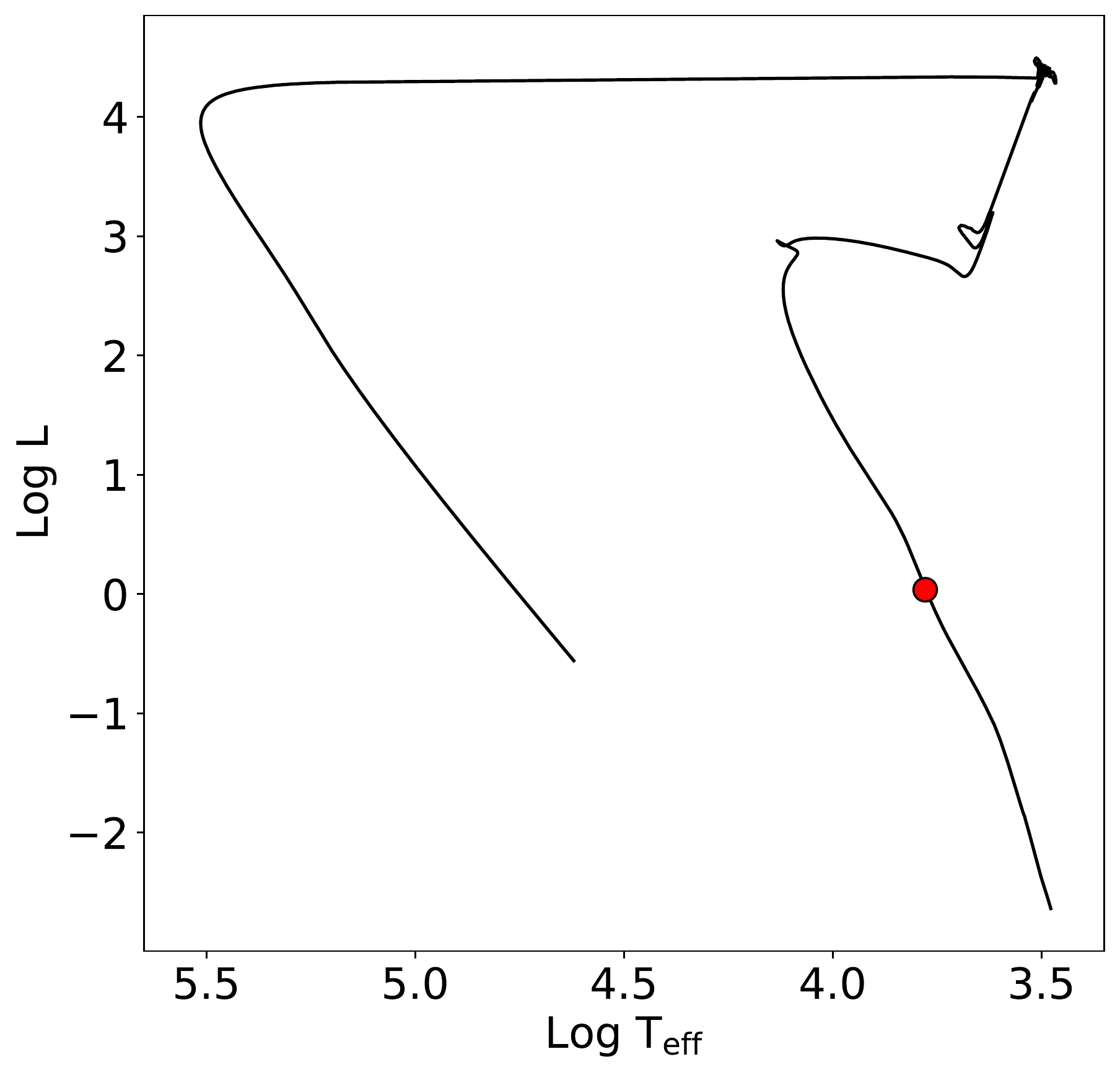}
    \includegraphics[scale=0.45]{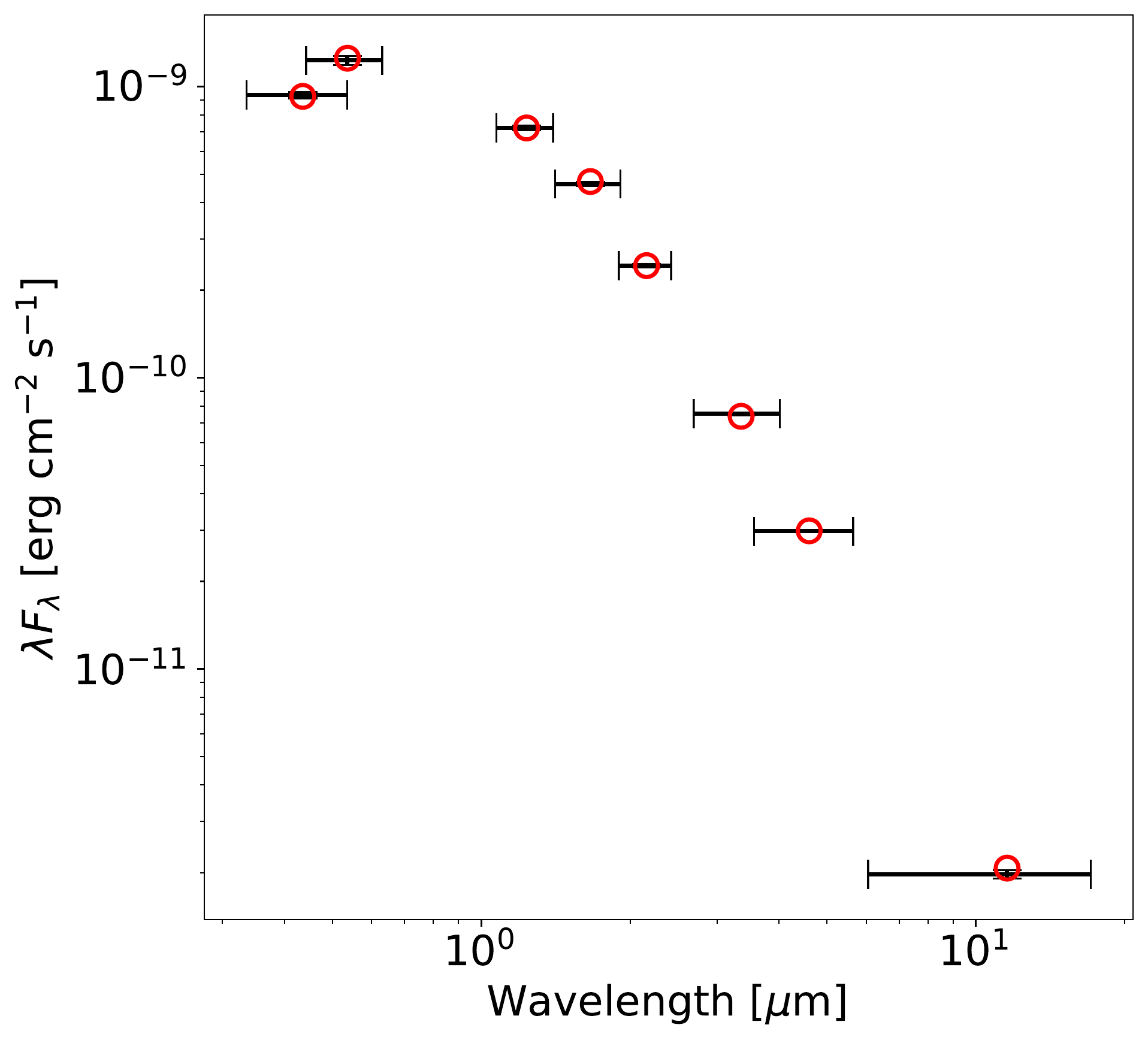}
    \caption{Isochrone (left) and SED (right) fits for Gaia DR2 69847610026687104, a Pleiades member with $T_{\rm eff} \sim 6000$ K and $R \sim 0.96$ \rsun. Red markers indicate the model fit to the black isochrone and stellar magnitudes. The global log-evidence for this fit is similar to the mean log-evidence for the Pleiades sample. The wavelength and reference flux values used in the right panel come from the SVO Filter Profile Service \citep[][]{svofilters1, svofilters2}.}
    \label{fig:isochrone_sed_fitting}
\end{figure*}

We used the \texttt{zero-point} code \citep{lindegren2020_gaia_edr3_parallax_bias} to estimate and correct the zero-point parallax bias of each measurement. We set Gaussian priors on each cluster's age, metallicity, and reddening. We used Bayestar19 3D dust maps \citep[][]{green2019} for $V$-band reddening estimates, LAMOST DR6 \citep[][]{cui2012, hamer2021_lamost_data} for metallicity, and multiple sources to set age priors. Table \ref{tbl:priors} details the cluster-level priors and their sources. In addition to those priors, we used the (corrected) Gaia parallaxes and uncertainties for individual stellar distance priors. We then ran nested sampling with \texttt{multinest} \citep[][]{feroz2009_multinest} to determine the stellar parameters. Figure \ref{fig:isochrone_sed_fitting} shows the best-fitting isochrone and flux values for Gaia DR2 69847610026687104, a Pleiades member with $T_{\rm eff} \sim 6000$ K, illustrating the typical quality of fit.

\begin{table*}
\begin{center}
\caption{Gaussian prior values for isochrone fitting and their sources.}
\label{tbl:priors}
\begin{tabular}{cccc}
\hline
\hline
Cluster & Parameter & Value & Source \\
\hline
  & $\log{\rm{age}}$ & $8.13 \pm 0.08$ & \citet[][]{gossage2018} \\
 Pleiades & [Fe/H] & $-0.05 \pm 0.18$ & \citet[][]{cui2012,hamer2021_lamost_data} \\
  & $A_{V}$ & $0.27 \pm 0.29$ & \citet[][]{green2019} \\
  \hline
  & $\log{\rm{age}}$ & $8.87 \pm 0.05$ & \citet[][]{brandt2015, gossage2018} \\
 Praesepe & [Fe/H] & $0.14 \pm 0.17$ & \citet[][]{cui2012,hamer2021_lamost_data} \\
  & $A_{V}$ & $0.013 \pm 0.039$ & \citet[][]{green2019} \\
  \hline
  & $\log{\rm{age}}$ & $8.18 \pm 0.08$ & \citet[][]{meibom2009} \\
 M35 & [Fe/H] & $0.08 \pm 0.15$ & \citet[][]{cui2012,hamer2021_lamost_data} \\
  & $A_{V}$ & $0.69 \pm 0.15$ & \citet[][]{green2019} \\
\hline
\end{tabular}
\end{center}
\end{table*}

\subsection{Inclination determinations}
\label{subsec:inclination_measurements}
\begin{figure}
    \includegraphics[scale=0.45]{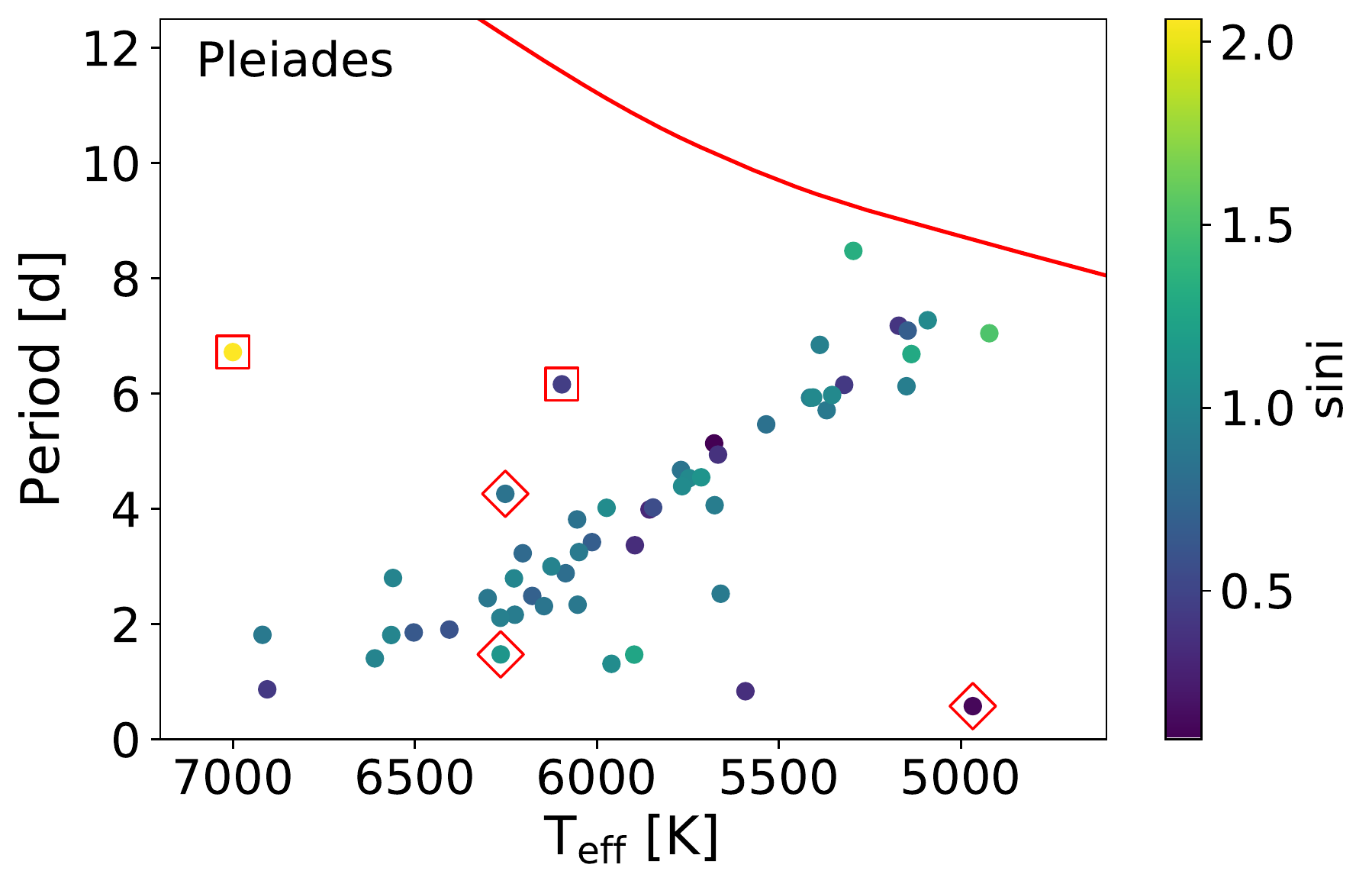}
    \includegraphics[scale=0.45]{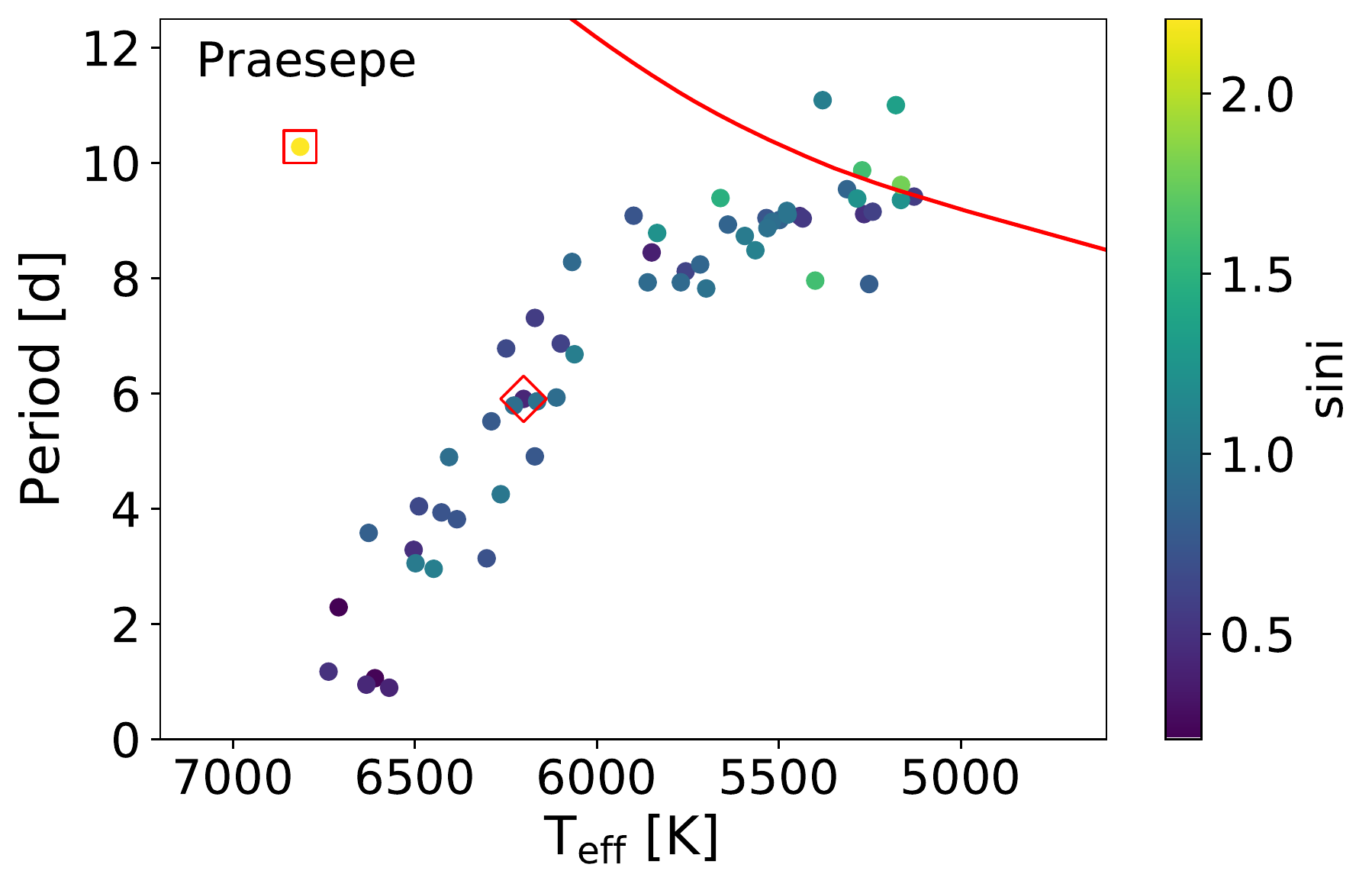}
    \includegraphics[scale=0.45]{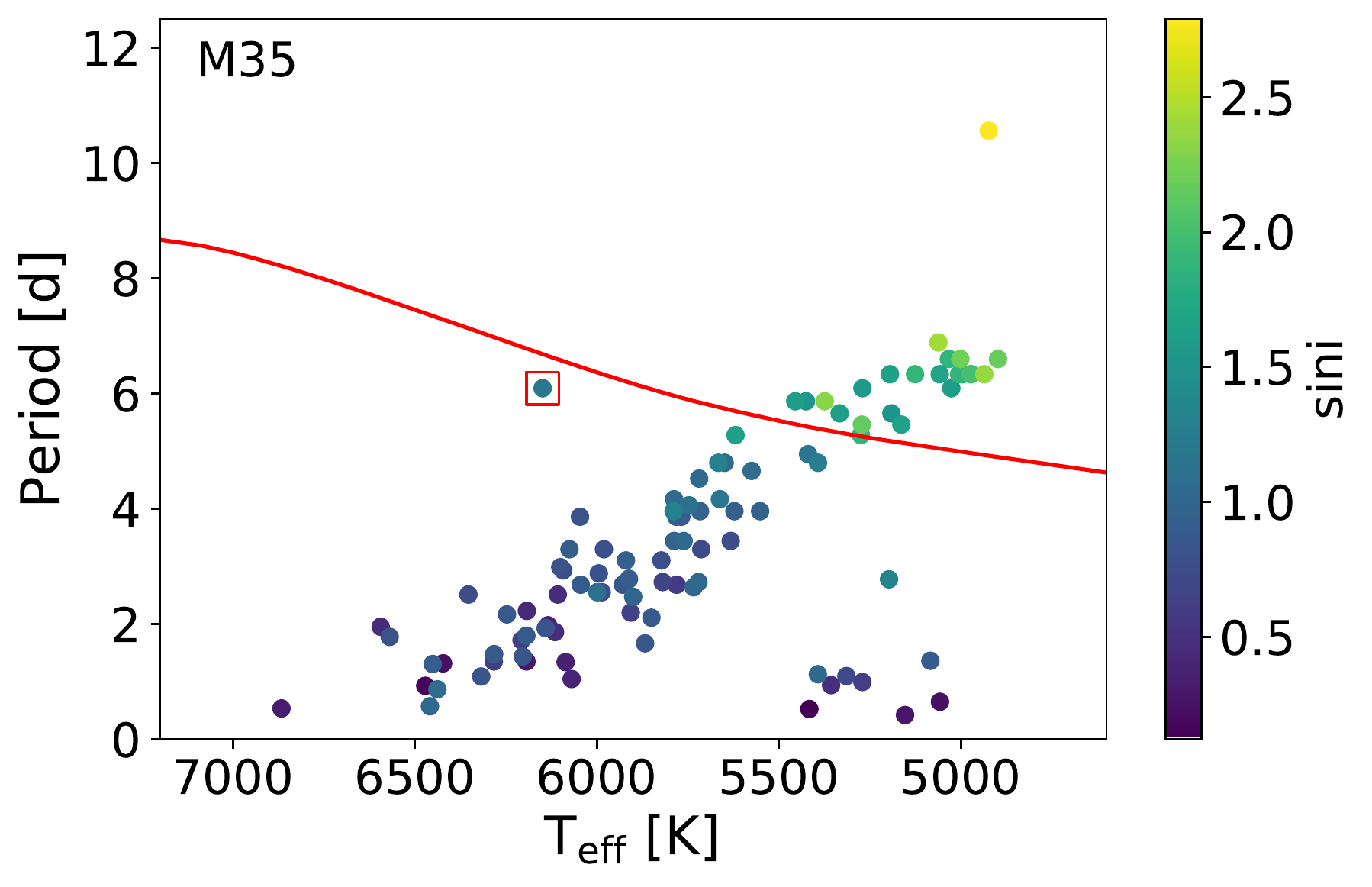}
    \caption{Period--effective temperature plots showing measurements for each cluster that are color-coded by the preliminary $\sin i$ value they yield. Determinations of $\sin i$ greater than unity are possible due to the combined influence of random and systematic errors on the measurements that yield the inclination.
    Above the red contours, stars are expected to have $v\sin i$ less than the applied thresholds (5 km s$^{-1}$ for Pleiades and Praesepe; 10 km s$^{-1}$ for M35), suggesting that their reported $v\sin i$ values are spurious. We removed these stars from the sample.
    Diamonds highlight stars removed due to line-of-sight velocities that are significantly discrepant with the cluster's distribution, while boxes indicate other excluded stars with rotation periods that are highly discrepant with empirical predictions.}
    \label{fig:per_teff_plots}
\end{figure}

We combined $v\sin i$, period, and radius measurements to yield the sine of the inclination of each target star (\ref{eq:spectrophot}). In an effort to minimize the effect of systematic measurement errors on our results, we discarded $v\sin i$ values in cases where our simulations predicted no measurements above a cluster's threshold of $v\sin i$. We also removed radial velocity outliers to exclude additional likely binaries and also excluded stars whose rotation periods were highly discrepant from color--period predictions \citep[e.g.][]{barnes2003} to avoid potentially inaccurate period values. The younger Pleiades and M35 clusters display both fast- and slow-rotating sequences, while Praesepe only shows the latter. Overall, this process removed 5 stars from the Pleiades sample, 6 from Praesepe, and 24 from M35. The red boxes and diamonds in Figure \ref{fig:per_teff_plots} visualize this selection on plots of period versus effective temperature.

For the Pleiades and Praesepe, we included only stars with $v\sin i > 5$ km s$^{-1}$, respectively removing 6 and 18 stars from these clusters' samples. Rotational velocities lower than this threshold are too similar to the stellar macroturbulent velocity \citep[e.g.][]{doyle2014}. These two sources of broadening are very difficult to disentangle, even at high resolution. While including smaller values of $v\sin i$ would increase our sample size, the values may not accurately quantify the rotation velocity and could bias the inclination distribution. The WOCS study reported $v\sin i $ only for values greater than 10 km s$^{-1}$, so we used that threshold for M35. We further discuss macroturbulence in Sec. \ref{subsec:possible_systematics}.

We also identified potential systematic errors in $v\sin i$ by calculating the maximum expected $v\sin i$ value (corresponding to an edge-on inclination) across a grid of effective temperature and period values. We used the MIST isochrone corresponding to each cluster's priors for SED fitting (Table \ref{tbl:priors}) to relate effective temperature to radius values. We incorporated the empirical scatter seen in our sample of period and radius values, and for each cluster we drew a contour line (red curves in Figure \ref{fig:per_teff_plots}) representing the combinations of $P$ and $T_{\rm eff}$ yielding a maximum $v\sin i$ below our threshold. The stars above this contour potentially suffer from a systematic error that leads to a higher $v\sin i$ measurement than possible given the star's rotation period and radius. We discarded these measurements due to the sensitivity of the spectro-photometric method to systematic errors \citep[e.g.][]{kamiaka2018}. These contours assume a combination of scatter in $R$ and $P$ that yields greater-than-average $v\sin i$ values in order to avoid removing more stars than justified. The points in Figure \ref{fig:per_teff_plots} are color-coded by their respective $\sin i$ value. Previous studies, e.g. by \citet[][]{kovacs2018}, eliminated stars with outlying $\sin i$ values greater than a given threshold (e.g. 2.0), and our method achieves a similar result of removing unphysical inclination values using a more data-driven approach.

After the selection process, we calculated $\sin i$ for each star using a Bayesian analysis of the spectro-photometric measurements (as in \paperone, following \citet[][]{masuda2020}).

\subsection{Modeling the inclination distribution}
\label{subsec:twothreeparam_models}
In order to forward-model the selection effects induced by the $v\sin i$ threshold described in the previous section, we generated a Monte Carlo sample of simulated stars for each cluster. We used Gaussian kernel density estimation to approximate the de-reddened Gaia $BP - RP$ color distribution for each cluster. In the Pleiades and M35, we separated the stars into the fast- and slow-rotating sequences before determining their color distribution. We then generated $10^5$ simulated stars using the density estimation for slow-rotating stars in each cluster, and we supplemented the Pleiades and M35 samples with a proportionate amount of fast rotators.

For each of these stars, we used a MIST isochrone to relate the color to an effective temperature, radius, and mass. We added scatter to the color--radius relation consistent with the observed dispersion of each sample. We then assigned a rotation period to each star based on polynomial spline fits to the color--period data, again adding appropriate empirical scatter. We calculated the true rotation speed $v$ for each star using the de-projected form of Eq. \ref{eq:spectrophot}.

To convert this speed into the projected rotation velocity $v\sin i$, we generate projected inclination values using the two-parameter model first described by \citet[][]{jackson2010}. This model assumes that all stellar spins in a cluster are uniformly distributed within a cone having a mean inclination $\alpha$ and spread half-angle $\lambda$. The average projected orientation of cluster members' spin axes is reflected in $\alpha$, which can vary from $0^{\circ}$ (pole-on) to $90^{\circ}$ (edge-on). The degree of alignment among the spin axes is quantified by $\lambda$, for which values close to the minimum of $0^{\circ}$ represent tight alignment, while a maximal half-angle value of $90^{\circ}$ corresponds to total isotropy. Since the spectro-photometric method does not permit a constraint on the azimuthal orientation of stellar spin axes in the plane of the sky, the above two parameters parameterize the model's projected inclination distribution.

Each model CDF is generated using a Monte Carlo method. We pair each simulated $v$ to a $\sin i$ value, removing stars with $v\sin i$ below the corresponding threshold for that cluster. An increasingly strict $v\sin i$ threshold will remove greater fractions of slower-rotating and low-inclination stars from a sample.

Finally, we simulate the effect of measurement uncertainties on the inclination distribution by assuming a Gaussian fractional error distribution for the rotation period and radius, and a log-normal distribution for fractional error in $v\sin i$. This latter choice approximates well the v sin i data in our study and was also used by \citet[][]{kovacs2018}. We determine the parameters used to generate the simulated observational errors by calculating the statistics corresponding to the Gaussian and log-normal distributions on the data for each cluster.

\subsection{Determining inclination model parameters}
\label{subsec:determining_inclination_parameters}
In \paperone, we optimized the model parameters $\alpha$ and $\lambda$ by mapping their possible values to ordered pairs spaced in 1$^{\circ}$ intervals. This imposition of Cartesian coordinates assigns equal weighting to each ordered pair, potentially favoring near-isotropic solutions ($\lambda \sim 90^{\circ}$) because of the similarity of the resulting inclination distribution even for many different values of $\alpha$.
In this paper, we performed a 10,000-step, 50-walker Markov Chain Monte Carlo (MCMC) analysis, setting uniform priors for $\alpha$ and $\lambda$ that covered the full range of parameter space without forcing specific combinations of the parameters. We again defined a Gaussian likelihood function (LF) based on the reduced $\chi^2$ calculated from each model and empirical CDF (Eq. \ref{eq:likelihood}):

\begin{equation}
    \ln{\rm{LF}} = -0.5 * \sum_{i=1}^{\rm{n}} \bigg[\ln{(2\pi \sigma_{i}^2)} + \frac{(x_{i} - \hat{x_{i}})^2}{\nu \sigma_{i}^2} \bigg].
    \label{eq:likelihood}
\end{equation}
Above, $\rm{n}$ is the number of $\sin i$ values for a cluster, $x_{i}$ is each determined $\sin i$ value, $\sigma_{i}$ is the uncertainty in $\sin i$, and $\hat{x_{i}}$ values are the model's $\sin i$ values interpolated onto the array of cumulative probabilities for the data. Finally, $\nu = \rm{n} - \rm{m}$ represents the degrees of freedom used to calculate the reduced $\chi^2$, with the number of fitted parameters $\rm{m} = 2$ ($\alpha$ and $\lambda$). After discarding a burn-in of 1000 steps, we marginalized the posteriors to get probability distributions (PPDs) for $\alpha$ and $\lambda$.

To gauge the performance of our MCMC analysis method, we validated it on four simulated inclination distributions. The first three contained 30 stars and the last contained 120. The first simulated distribution was isotropic, the second was tightly aligned with parameters $\alpha = 30^{\circ}$ and $\lambda = 25^{\circ}$, and the final two shared the same moderate alignment ($\alpha = 75^{\circ}$ and $\lambda = 45^{\circ}$) with two sample sizes (30 and 120 stars). To model the measurement uncertainties for the first two validation runs, we used the Pleiades sample's error distributions. We also use our simulated true rotation velocity distribution for the Pleiades to model a 5 km s$^{-1}$ $v\sin i$ threshold. For the second two runs, we used the Praesepe sample's error distributions and simulated rotation velocities. In these latter two runs, we applied a 5 km s$^{-1}$ $v\sin i$ threshold to the simulated 30 star sample, and no threshold to the 120 star sample.

The first simulation yielded the expected degeneracy between isotropic and moderately aligned parameters, with a PPD favoring $\lambda = 89^{\circ}$ at its peak with a median value of $\sim 79^{\circ}$. The 95\% confidence interval for $\lambda$ extends as low as $\sim 48^{\circ}$, while it constrained $\alpha$ between 21$^{\circ}$ and 85$^{\circ}$. The second, tightly aligned test put an upper limit on $\alpha \lesssim 31^{\circ}$ and also constrained $\lambda$ to between 20$^{\circ}$ and 49$^{\circ}$ at 95\% confidence. This uncertainty in the spread half-angle reflects another known degeneracy, this time between a tightly aligned spread angle at larger mean inclinations and a larger spread angle with a smaller mean inclination.

The third, moderately aligned 30 star simulation did not yield a statistically significant departure from isotropy, finding the most probable $\lambda = 83^{\circ}$ and $\lambda > 8^{\circ}$ at 95\% confidence. This test also recovered $\alpha = 55^{\circ}$ while allowing it to range between 9$^{\circ}$ and 85$^{\circ}$ at 95\% confidence. The fourth test, having the same distribution as the third but four times as many stars (and no $v\sin i$ threshold), yielded results similar to the third test: the peak $\lambda = 89^{\circ}$ and $\lambda > 7^{\circ}$ at 95\% confidence, while $\alpha = 63^{\circ}$ was constrained to values between $19^{\circ}$ and $87^{\circ}$. While the input moderately aligned parameters are not excluded by these results, this distribution cannot be significantly distinguished from isotropy due to the similarity of the two scenarios.

In these simulations, the increase from 30 to 120 stars did not yield a commensurate improvement in the accuracy or precision of the spread half-angle and mean inclination, highlighting the diminishing returns of analyzing a sample of $\sim 10^2$ $\sin i$ values versus $\sim 10^1$. Overall, the tests demonstrate our model's ability to distinguish between isotopic and tightly aligned spins aimed along the LOS, but they highlight the difficulty of breaking the degeneracy in parameters describing cases of isotropy and moderate alignment, even when analyzing a larger sample of stars with a looser quality cut.

We also experimented with adding a degree of freedom to the two-parameter model. The three-parameter model contains the parameter $f$, which describes the fraction of anisotropic stellar spins in a cluster. The remaining fraction $(1 - f)$ of stars is assumed to be isotropically oriented. We labeled the original two parameters as $\alpha_{\rm aniso}$ and $\lambda_{\rm aniso}$ in this model to signify that they now describe the aligned stars, rather than the entire cluster. The three-parameter model can be employed when a two-parameter fit provides unsatisfactory results. Its use is especially supported when additional data suggest the possibility of an anisotropic subset of spins in a cluster.

\subsection{Cluster kinematics}
\label{subsec:kinematics_analysis}
To facilitate a comparison between the inclinations and overall motion of cluster members, we performed an analysis of astrometric and spectroscopic data in a similar manner to \paperone. The differences from that paper's analysis are the use of Gaia EDR3 proper motions and the correction of these measurements for biases in bright ($G < 13$) stars via Eq. 5 of \citet[][]{cantatgaudin2021_pm_correction}.

For each star meeting the photometric, classification, and RUWE criteria described in Sec. \ref{subsec:member_selection}, we used the technique of \citet[][]{kamann2019} to determine the plane-of-sky (tangential/radial) and LOS kinematics as a function of radial distance from cluster center. We transform Gaia proper motions (with mean cluster motion subtracted) into polar coordinates with the origin at the center of the cluster to obtain plane-of-sky kinematics. Tangential motion corresponds to the azimuthal direction in this coordinate system. LOS velocities, directed perpendicular to the plane of the sky, are measured using the Doppler shift of stellar spectral lines.\footnote{LOS velocities are often called ``radial velocities'' in the literature, but we exclusively use ``LOS'' in this work to avoid confusion with the radial component of proper motions and their associated velocities.}

We also use the notation of \citeauthor{kamann2019} to label our results: in all three directions, $v_{0}$ is the mean velocity of stars and $\sigma$ is the velocity dispersion. In the LOS direction, we also consider the position angle $\theta_{c}$ and rotation velocity $v_{\rm rot}$. Due to the smaller number of LOS velocities compared with proper motions, we assume that $v_{\rm rot}$ and $\sigma_{\rm LOS}$ are constant.
 
We again used Eq. 2 of \citet[][]{gaiaclusters2018} to map R.A. and decl. to a Cartesian coordinate system, Eq. 6 of \citet[][]{vandeven2006} to calculate and subtract motion due to perspective effects (e.g. cluster motion along the LOS), and Eq. 10 of \citet[][]{vanleeuwen2000} to convert the resulting corrected proper motions to polar coordinates. Gaia EDR3 provided the LOS velocities for the Pleiades and Praesepe, while for M35 we used the WOCS study \citep[][]{m35_vsini_leiner_2015} for its greater number of measurements.

\section{Results} \label{sec:results}

\subsection{Isochrone/SED fitting}
We visually inspected the isochrone/SED fit results both individually and as an ensemble to verify that the (1) the data points were well fit and (2) stellar parameters such as age, [Fe/H], $A_{V}$ and distance agreed with the expected value for each cluster. For each star, we also examined the logarithm of the Bayesian evidence ($\ln{Z}$) determined by approximating an integral of the likelihood times the prior over the parameter space. The evidence can assist in selecting the appropriate hypothesis or model to describe the data, and \citet{feroz2019} further discuss \texttt{multinest's} estimation of this quantity.

The vast majority of stars we studied had comparable evidence values, but a small fraction of the values were significantly lower. These smallest evidence values corresponded to the poorest SED fits, and they may indicate the unsuitability of our hypothesis/model in these cases, for example due to a binary companion or non-member of a cluster. We discarded stars having significantly lower evidence values than the main grouping to avoid including inaccurate radius values in our inclination analysis. The fraction of stars that remained after this selection was 94\% for the Pleiades, 96\% for Praesepe, and 98\% for M35.

\subsection{Inclination distributions}
\begin{figure*}
    \centering
    \includegraphics[scale=0.45]{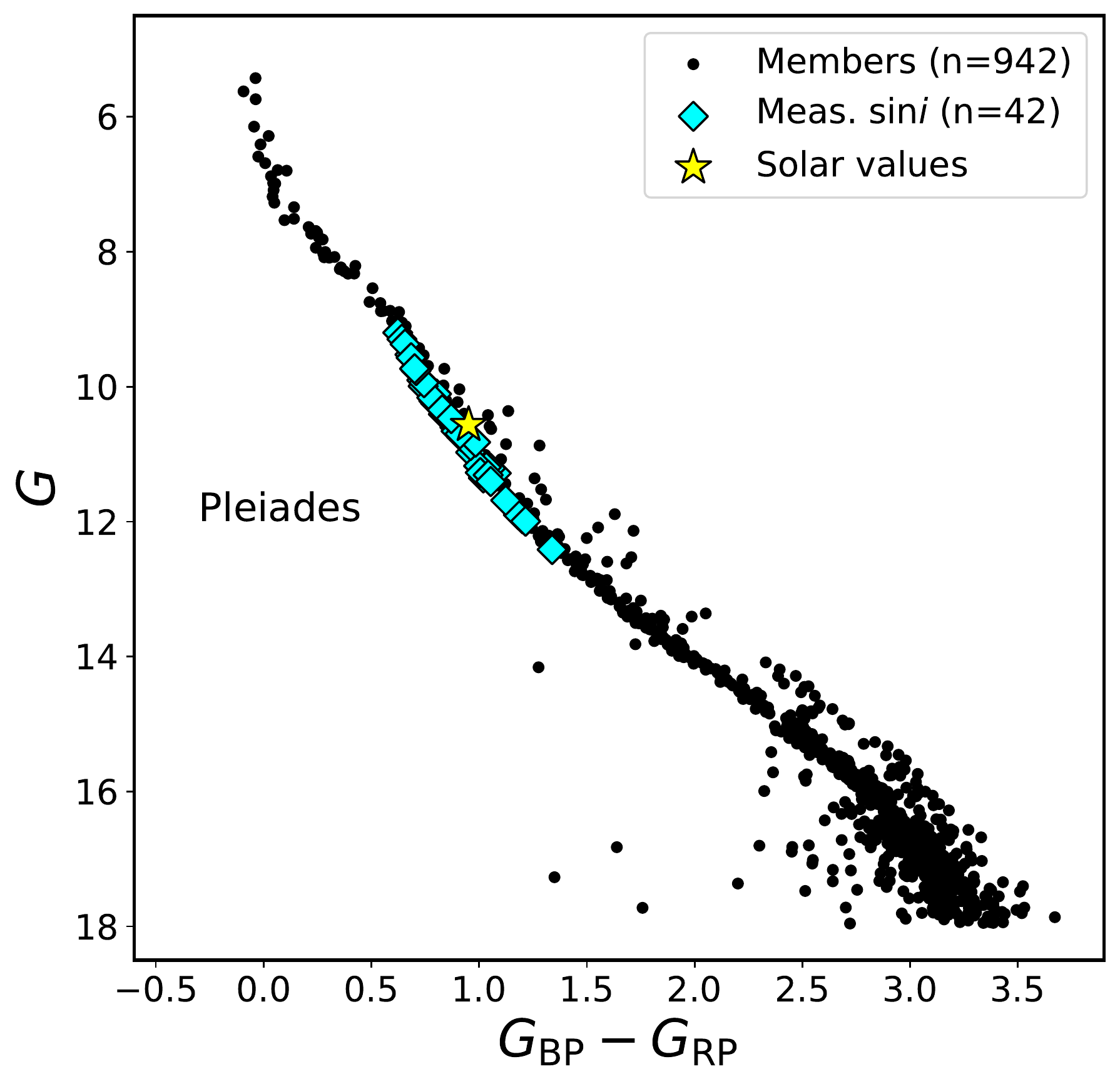}
    \includegraphics[scale=0.45]{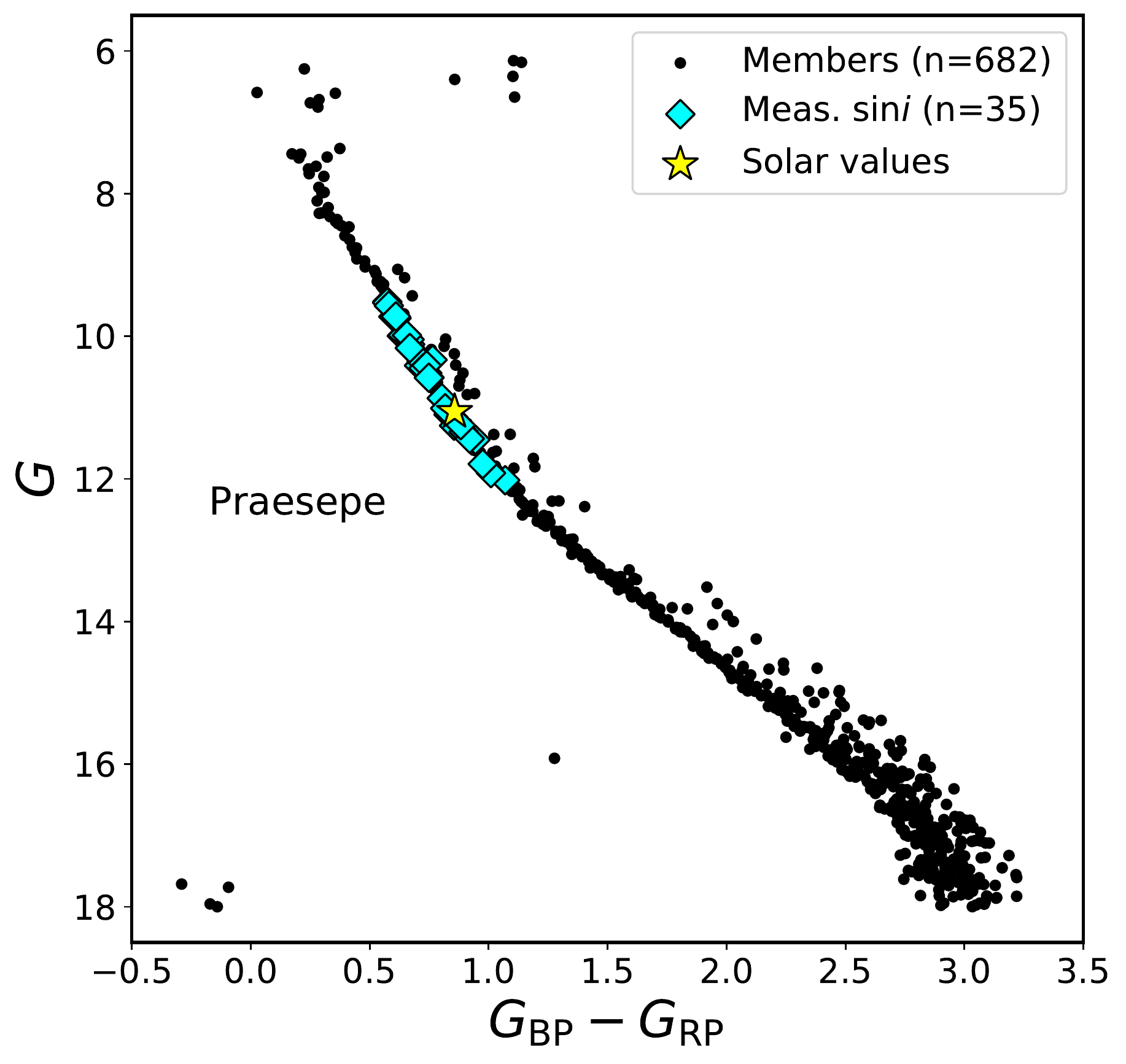}
    \includegraphics[scale=0.45]{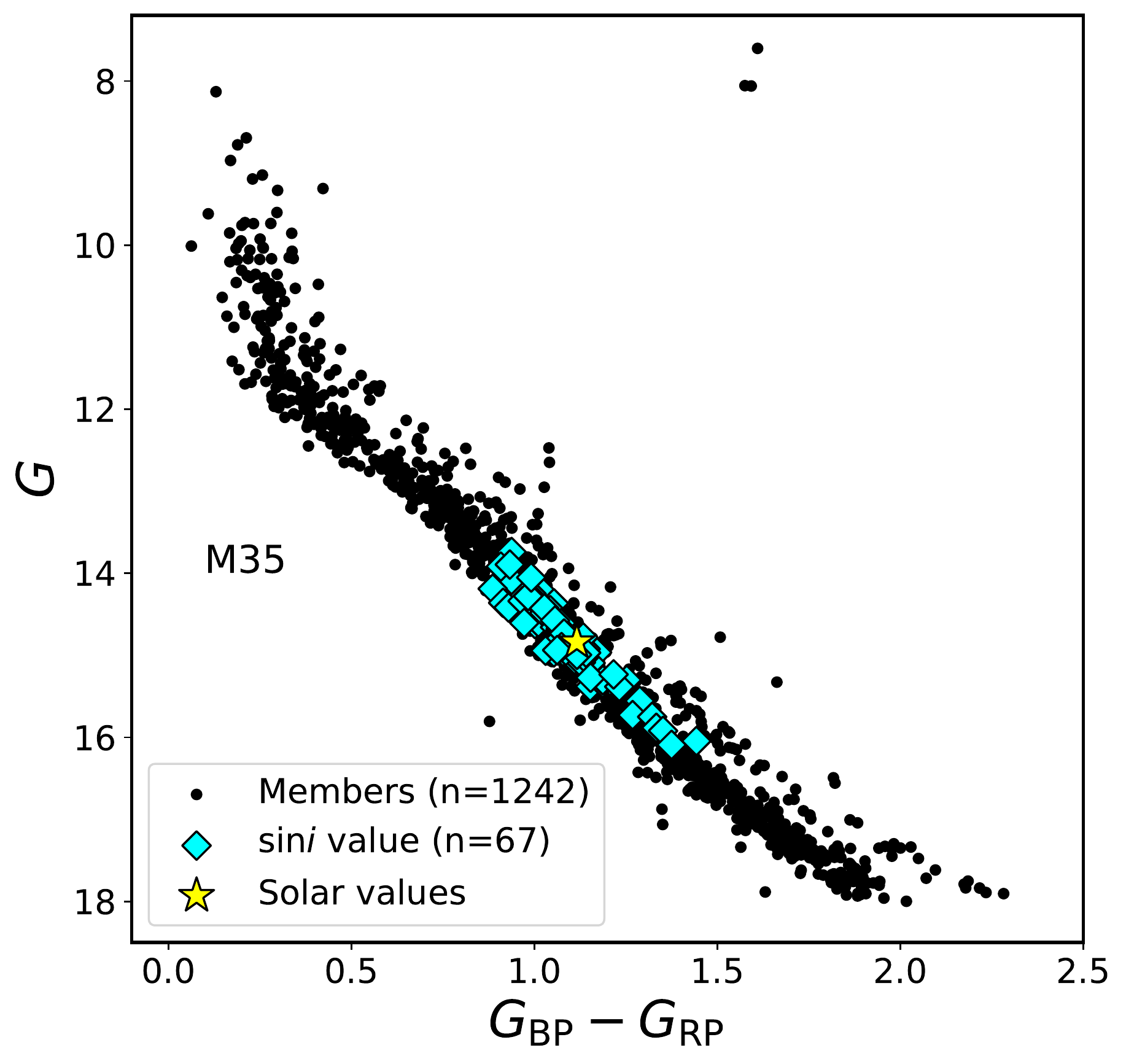}
    \caption{Color--magnitude diagrams of each cluster's Gaia DR2-based members \citep[][]{cantatgaudin2018} with probability $> 0.68$ (black) and stars with $\sin i$ determined in this work (cyan). We plot the Gaia DR2 colors and magnitudes for consistency with the original member list.
    Estimated values for the Sun at the distance and reddening of each cluster are marked by the yellow star.}
    \label{fig:cmd_plots}
\end{figure*}

\begin{figure*}
    \centering
    \includegraphics[scale=0.425]{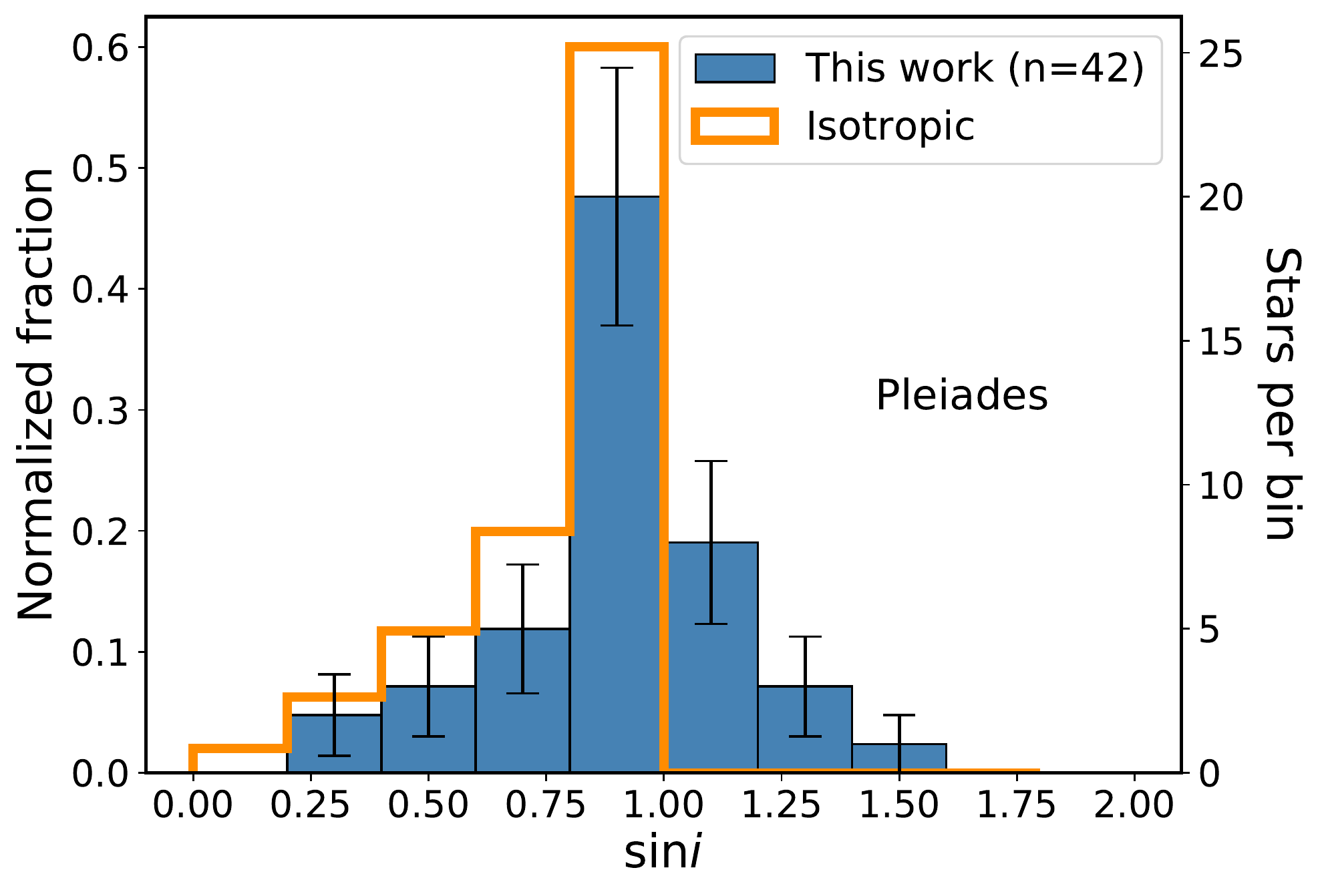}
    \includegraphics[scale=0.425]{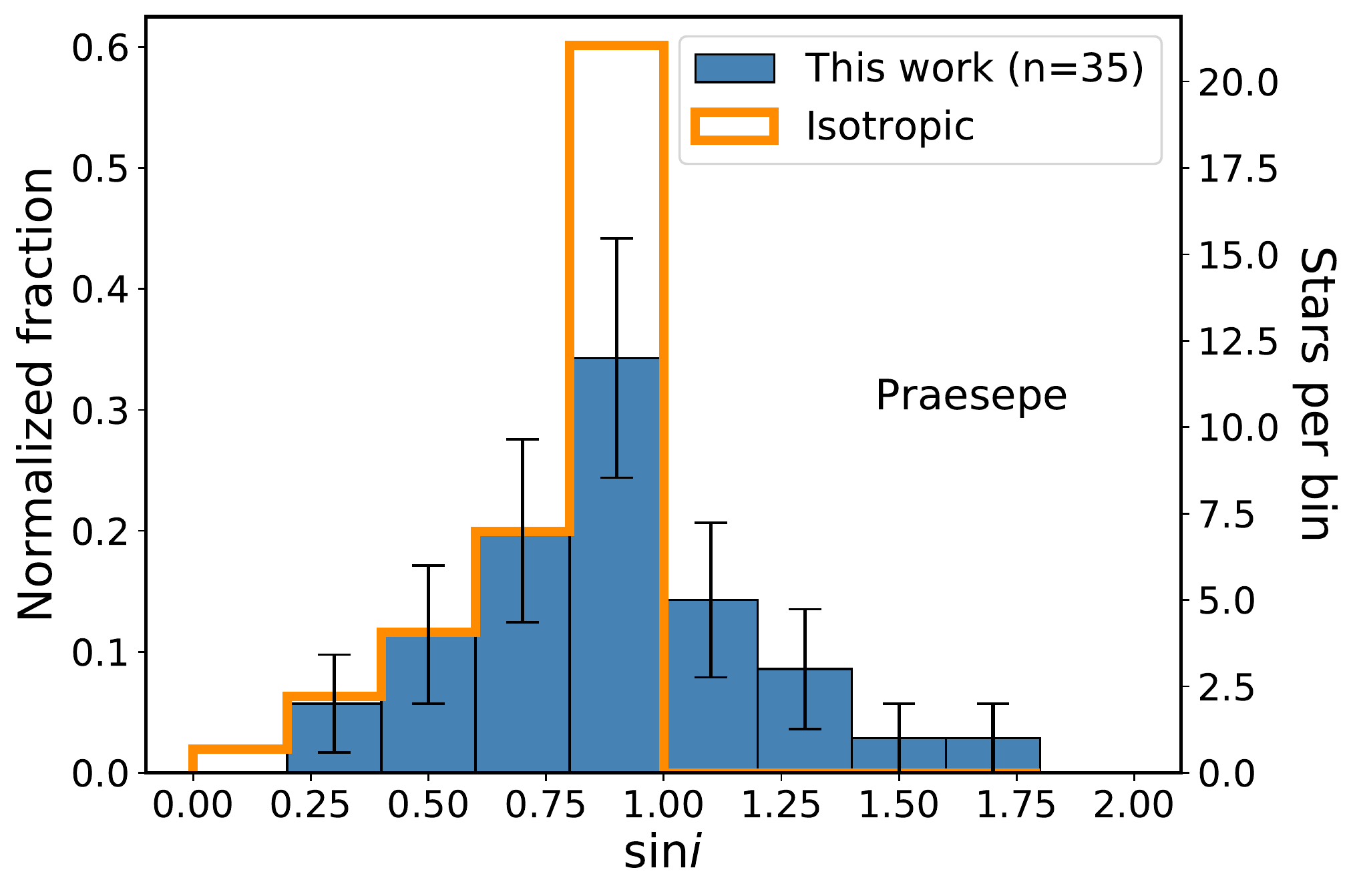}
    \includegraphics[scale=0.425]{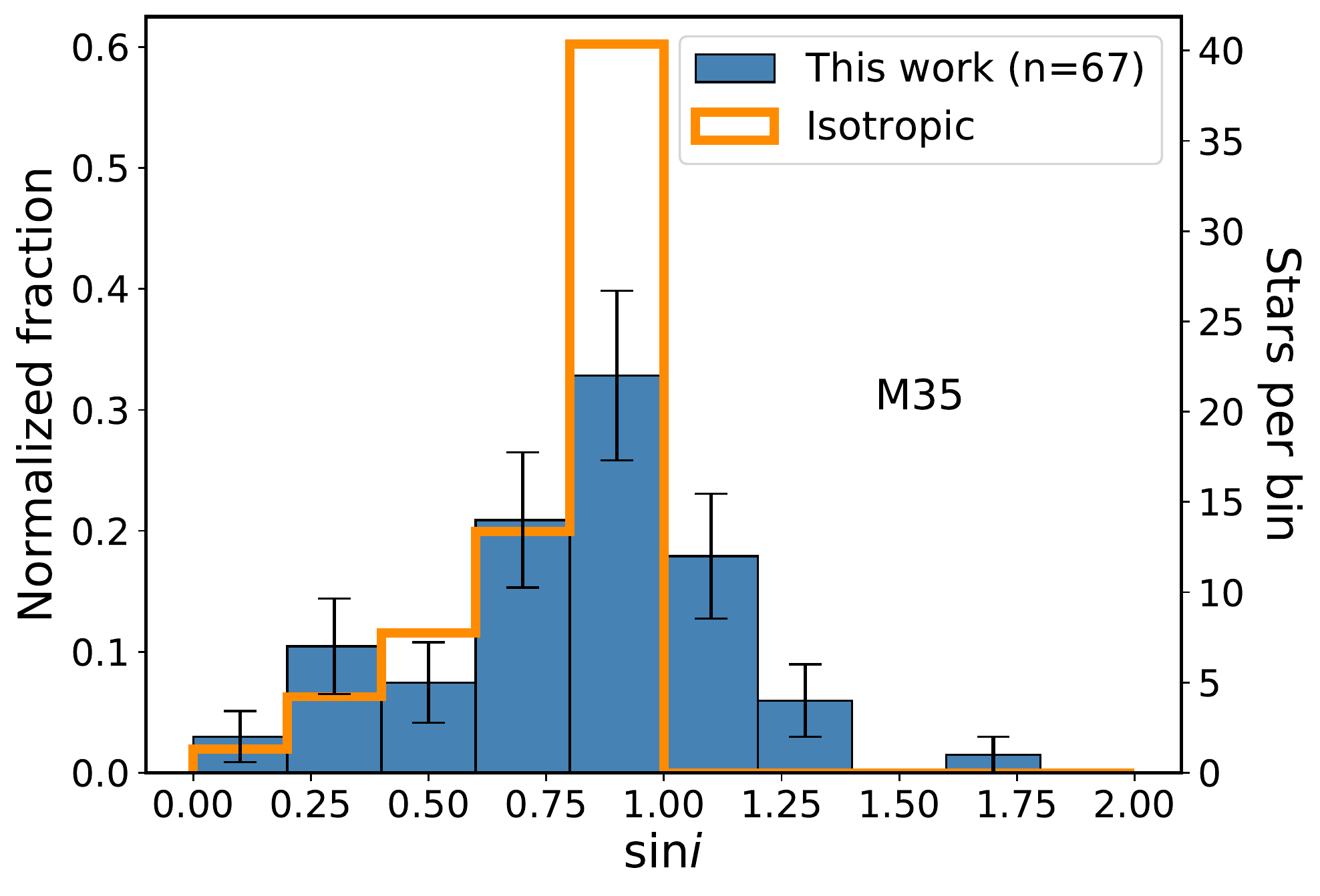}
    \caption{Histograms of this work's $\sin i$ determinations for each cluster, with Poisson error bars plotted for each bin. Orange lines indicate the theoretical isotropic distribution (uniform in $\cos i$).}
    \label{fig:sini_histfigs}
\end{figure*}

After applying each cluster's $v\sin i$ threshold and removing spurious values, we have projected inclinations for 42 stars in the Pleiades, 35 in Praesepe, and 67 in M35. The mean uncertainty in $\sin i$ was 13\% for the Pleiades, 16\% for Praesepe, and 13\% for M35. Figure \ref{fig:cmd_plots} shows each cluster's color-magnitude diagram (CMD). Figure \ref{fig:sini_histfigs} shows histograms of each cluster's inclination distribution. Table \ref{tbl:results} reports each $\sin i$ value along with 16th and 84th percentile uncertainties.

\begin{table*}
\begin{center}
\caption{Relevant quantities and determinations of Pleiades, Praesepe and M35 inclinations.}
\label{tbl:results}
\begin{tabular}{ccccccccc}
\hline
\hline
GDR2 Source ID & R.A. [$^\circ$] & Decl. [$^\circ$] & $T_{\rm eff}$ [K] & Radius [\rsun] & $v\sin i$ [km s$^{-1}$] & Period [d] & $\sin i$ & Clust. \\
\hline
66809491897360896 & 56.73 & 24.8 & 5896 $\pm$\ 92 & 0.9818 $\pm$\ 0.0073 & 5.1 $\pm$\ 1.3 & 3.370 $\pm$\ 0.034 & 0.3441$^{+0.089}_{-0.089}$ & Pl \\
68303487680516224 & 56.11 & 24.59 & 5592 $\pm$\ 84 & 0.8706 $\pm$\ 0.0065 & 18.90 $\pm$\ 0.70 & 0.8374 $\pm$\ 0.0084 & 0.3571$^{+0.015}_{-0.016}$ & Pl \\
65225611037551360 & 56.66 & 24.1 & 6900 $\pm$\ 220 & 1.309 $\pm$\ 0.017 & 32.3 $\pm$\ 3.2 & 0.8702 $\pm$\ 0.0087 & 0.4221$^{+0.044}_{-0.044}$ & Pl \\
68322145018429952 & 56.06 & 24.78 & 5846 $\pm$\ 94 & 0.9449 $\pm$\ 0.0074 & 6.6 $\pm$\ 1.0 & 4.026 $\pm$\ 0.040 & 0.5522$^{+0.085}_{-0.086}$ & Pl \\
69811948914407168 & 56.17 & 24.82 & 6400 $\pm$\ 140 & 1.157 $\pm$\ 0.010 & 18.5 $\pm$\ 1.0 & 1.908 $\pm$\ 0.019 & 0.5992$^{+0.035}_{-0.036}$ & Pl \\
66960262430594816 & 57.57 & 25.38 & 6503 $\pm$\ 53 & 1.2433 $\pm$\ 0.0050 & 21.6 $\pm$\ 2.2 & 1.857 $\pm$\ 0.019 & 0.6332$^{+0.066}_{-0.067}$ & Pl \\
69847610026687104 & 56.61 & 25.14 & 6014 $\pm$\ 94 & 0.9612 $\pm$\ 0.0065 & 9.6 $\pm$\ 1.1 & 3.424 $\pm$\ 0.034 & 0.6722$^{+0.079}_{-0.079}$ & Pl \\
66523893753437824 & 57.11 & 24.05 & 6200 $\pm$\ 120 & 1.0926 $\pm$\ 0.0097 & 15.60 $\pm$\ 0.90 & 2.491 $\pm$\ 0.025 & 0.6982$^{+0.042}_{-0.044}$ & Pl \\
66788291938818304 & 56.59 & 24.57 & 6200 $\pm$\ 110 & 0.9866 $\pm$\ 0.0069 & 11.90 $\pm$\ 0.80 & 3.231 $\pm$\ 0.032 & 0.7653$^{+0.053}_{-0.054}$ & Pl \\
64941211186036352 & 57.14 & 23.43 & 6100 $\pm$\ 120 & 1.0103 $\pm$\ 0.0086 & 14.20 $\pm$\ 0.70 & 2.885 $\pm$\ 0.029 & 0.7963$^{+0.042}_{-0.043}$ & Pl \\
... & ... & ... & ... & ... & ... & ... & ... & ... \\
\hline
\end{tabular}
\end{center}
\tablecomments{Table \ref{tbl:results} is published in its entirety in the machine-readable format.}
\end{table*}

\begin{figure*}
    \centering
    \includegraphics[scale=0.4]{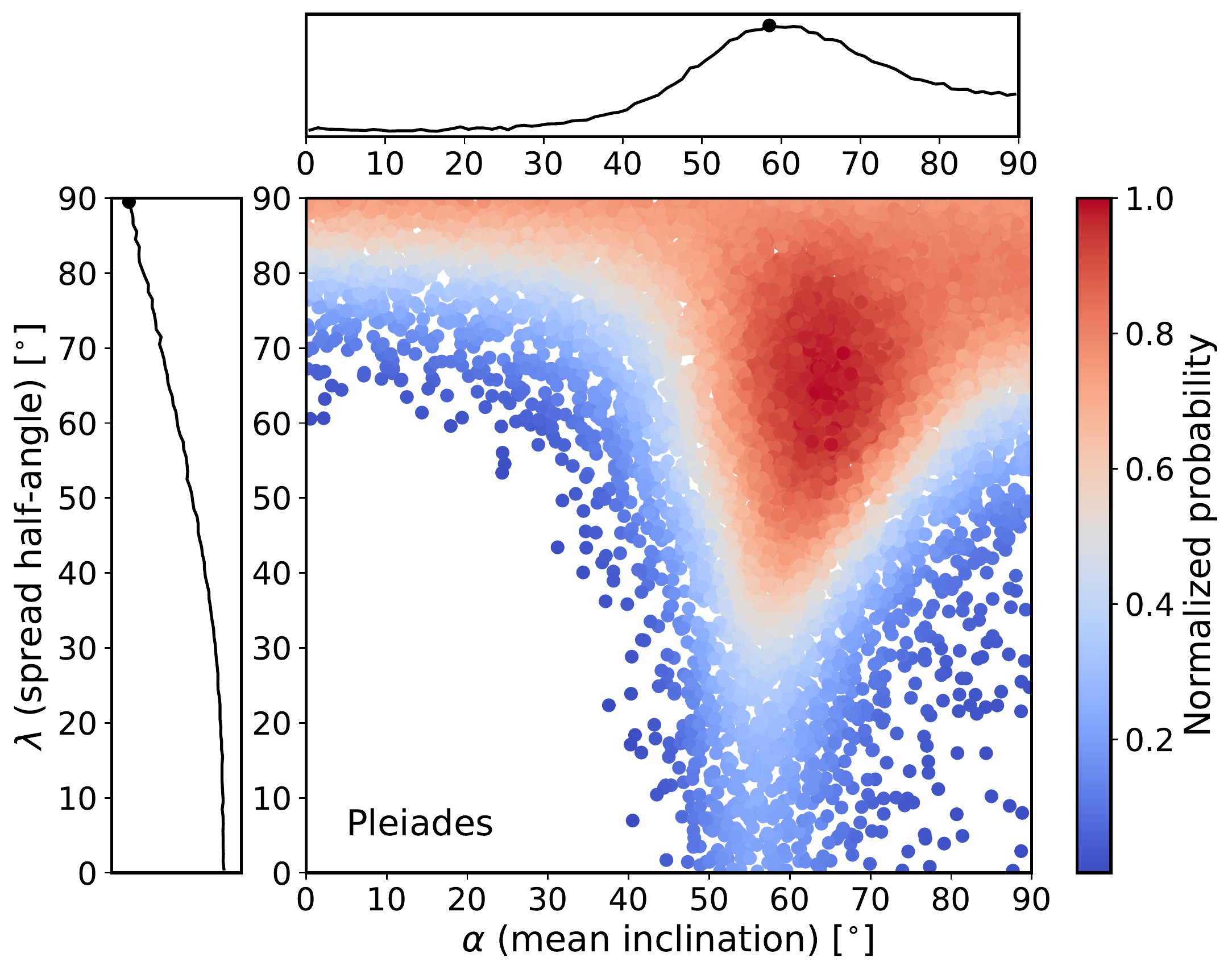}
    \includegraphics[scale=0.4]{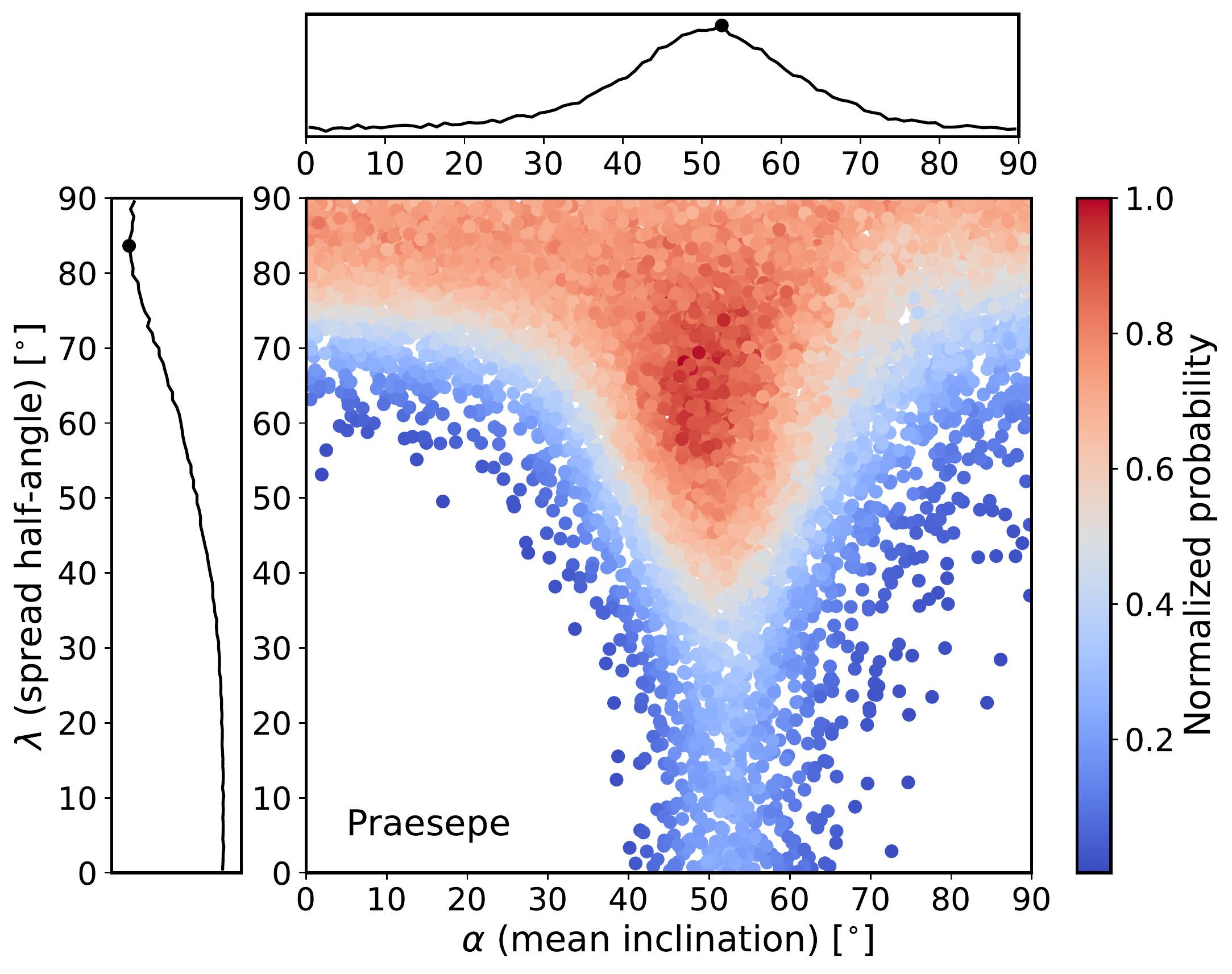}
    \includegraphics[scale=0.4]{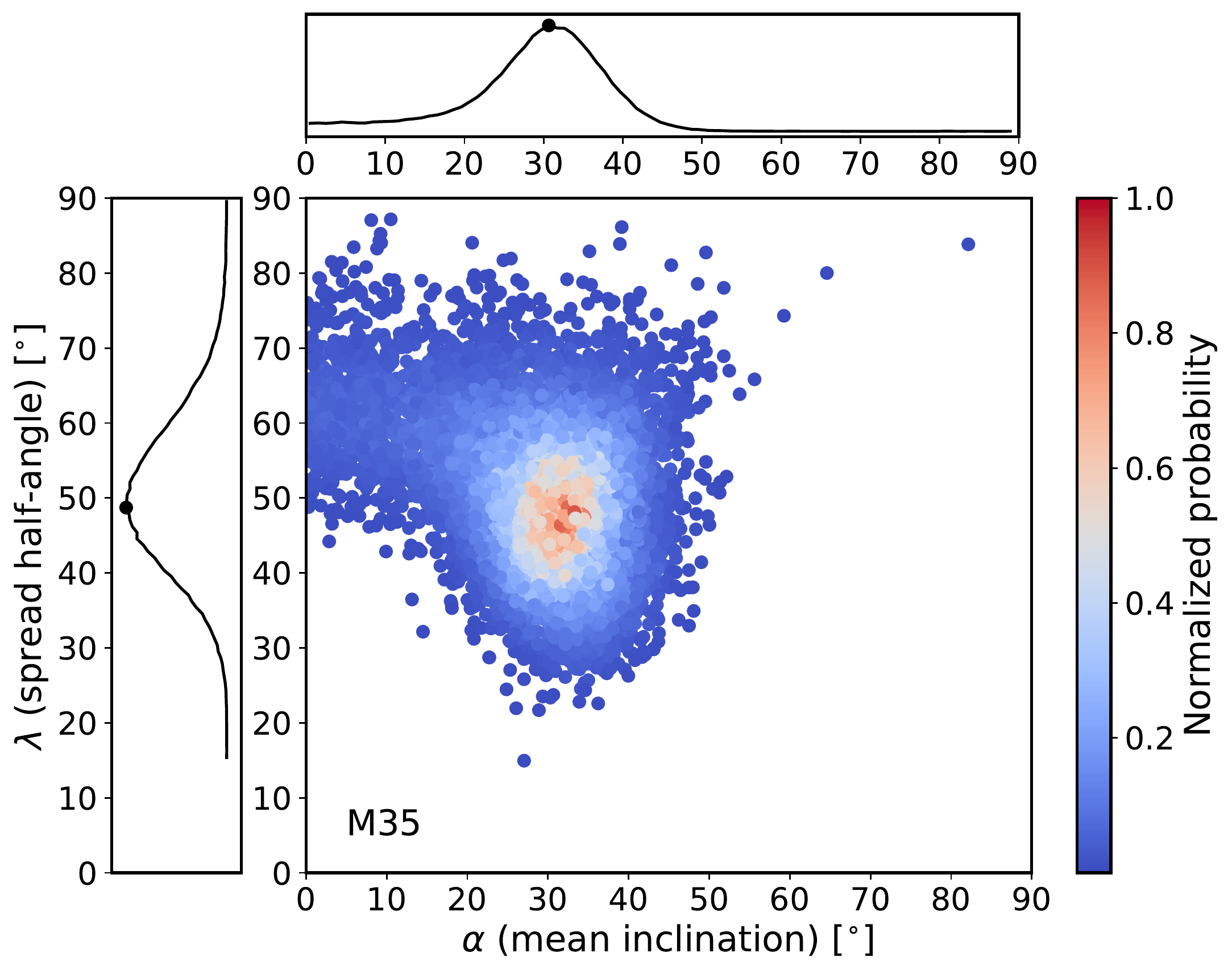}
    \caption{Plots of two-parameter model values and associated probabilities representing
    2\% of each cluster's MCMC chains. The model assumes that spin axes are uniformly distributed within a cone having mean inclination $\alpha$ and spread half-angle $\lambda$. We also plot the 1D marginal distributions for each parameter.}
    \label{fig:sini_chi2figs}
\end{figure*}

We show visualizations of two-parameter model probabilities across the full domain of $\alpha$ and $\lambda$ in Figure \ref{fig:sini_chi2figs}. Points on each plot show the combinations of parameters accepted by the MCMC run, with the entire plotted sample thinned to represent 2\% of all values in the chain.
The 95\% confidence intervals for the marginalized Pleiades parameters display tight to moderate alignment at moderate to high mean inclinations (peaking at $\lambda = 89^{\circ}$ and $\alpha = 60^{\circ}$, with $\lambda > 24^{\circ}$ and $13^{\circ} < \alpha < 85^{\circ}$).

The Praesepe spread half-angle PPD peaks at $\lambda = 85^{\circ}$ and $\alpha = 53^{\circ}$, constraining $\lambda > 26^{\circ}$ and $8^{\circ} < \alpha < 82^{\circ}$ at 95\% confidence. The Pleiades and Praesepe probability plots both show ``peninsula''-shaped model probabilities: while mostly surrounded by low-probability parameter combinations, there is a collection of parameters with high probability that extends from isotropic cases to moderate alignment at moderate mean inclinations. The similar probabilities among these different scenarios highlights a degeneracy in the parameters describing isotropic and moderately aligned spins for these clusters.

For the Pleiades and Praesepe, we also ran the MCMC analysis on the full datasets with no $v\sin i$ threshold. These runs yielded similar results to the analyses using the 5 km s$^{-1}$ threshold. The PPD peaks were at $\lambda = 89^{\circ}$, $\alpha = 60^{\circ}$ for the Pleiades and $\lambda = 88^{\circ}$, $\alpha = 55^{\circ}$ for Praesepe. At 95\% confidence for the Pleiades, we found $\lambda > 33^{\circ}$ and $12^{\circ} < \alpha < 85^{\circ}$. For Praesepe, we determined $\lambda > 6^{\circ}$ and $19^{\circ} < \alpha < 87^{\circ}$.

The third panel of Figure \ref{fig:sini_chi2figs} shows a distinct result for M35 compared to the previous two clusters. The M35 PPDs peak at $\lambda = 50^{\circ}$ and $\alpha = 32^{\circ}$. At 95\% confidence, we constrain this cluster's parameters to be $35^{\circ} < \lambda < 67^{\circ}$ and $10^{\circ} < \alpha < 41^{\circ}$. We plot CDFs of our $\sin i$ determinations for each cluster as black points in Figure \ref{fig:ecdf_figs} to allow further visualization of our results. Solid red curves display the model CDF for each cluster corresponding to the peaks of the PPDs determined with the two-parameter model, while dashed red curves show the isotropic distribution when it is not already plotted as the favored model. Pink shading represents parameters drawn from the 68\% confidence level distributions of $\alpha$ and $\lambda$. For M35, no two-parameter model CDF fits the data as well as the results for the Pleiades and Praesepe. The reduced $\chi^2$ value for the PPD-peak M35 model is $\sim 5$, while the corresponding values for the Pleiades and Praesepe are $\sim 1.2$ and $1.3$, respectively. While the isotropic model closely matches the data for $\sin i \gtrsim 0.75$ in M35, the inclinations smaller than this value are not well modeled.

\subsection{Grouping of Low-sini stars in M35}
We performed further analysis on the M35 inclination distribution due to the two-parameter model's underestimation of the occurrence of low-$\sin i$ ($\lesssim 0.75$) stars in the cluster. We employed the three-parameter model to provide further insight into this grouping of stars. Figure \ref{fig:m35_ecdf_twomodels} shows the best-fitting three-parameter model CDF to the M35 inclinations and compares it to the isotropic two-parameter distribution. The most probable values of the three-parameter model are $f \sim 0.1$, $\alpha_{\rm aniso} \sim 5^{\circ}$ and $\lambda_{\rm aniso} = 14^{\circ}$, yielding a reduced $\chi^2$ value of $\sim 0.3$.

We identified a subset of six stars in M35 having $\sin i < 0.35$ and corresponding most closely with the anisotropic fraction of the best-fitting three-parameter model. These stars are colored blue in Figure \ref{fig:m35_ecdf_twomodels}.
We explored the plane-of-sky spatial distribution of M35 stars in our sample, splitting the sample of 67 into two groupings on each side of the $\sin i = 0.35$ boundary. We plot the R.A. and decl. positions of all cluster members and these two subsets in Figure \ref{fig:m35_ra_dec_positions}.

We also studied the proper motions of the M35 stars. Using the same techniques as the kinematics analysis (Sec. \ref{subsec:kinematics_analysis}), we converted proper motions in the (R.A., decl.) directions to Cartesian coordinates ($x$, $y$) and subtracted the predicted motion due to the approach of M35 along the LOS. We plot the corrected proper motions for the sample of 67 stars in Figure \ref{fig:m35_corr_pm_x_y}, again splitting the data along the $\sin i = 0.35$ boundary. To compare the six low-inclination stars' proper motions to those of the remaining 61, we performed a bootstrapped, two-sample, two-dimensional K--S test \citep[][]{ndtest1, ndtest2, numerical_recipes_3}\footnote{\texttt{ndtest} Python code written by Zhaozhou Li, available at \url{https://github.com/syrte/ndtest}}
with 10,000 re-samplings, finding a $p$-value of 0.04.

To further examine the statistical significance of this six-star low-inclination grouping, we performed a Kuiper test, a version of the K--S test that is invariant to the phase of cyclic parameters, on the position angles (measured counterclockwise from north) of these stars. We compared these values to a uniform distribution of position angles, finding a $p$-value for compatibility with this null hypothesis of $\sim 0.12$. The magnitude of Gaia EDR3 parallax uncertainties for stars in this grouping prevent us from constraining their LOS positions relative to cluster center. We address potential sources of systematic error in Secs. \ref{subsec:possible_systematics} and \ref{subsec:diffrot}, and we discuss a physical interpretation of this subset in Sec. \ref{subsec:m35_physical_insights}.

\subsection{Cluster kinematics}
We determined tangential and radial kinematics using a sample of 749 stars in the Pleiades, 518 in Praesepe, and 959 in M35. In the LOS direction, we analyzed samples of 169, 144, and 246 stars, respectively. These stars passed the Gaia EDR3 RUWE selection as described in Sec. \ref{subsec:member_selection} in order to remove likely binaries. Table \ref{tbl:kinematics} lists each cluster's overall kinematic values. Uncertainties correspond to the 16th and 84th percentiles of each PPD. The top panels in Figure \ref{fig:cluster_rot_kinematics} plot the dependence on the binned tangential (blue) and radial (green) motions as a function of angular distance from each cluster center. Predicted radial motion due to each cluster's LOS velocity has been subtracted from each radial value. Each proper motion axis is accompanied by a conversion to physical units using an inversion of the mean of corrected EDR3 parallaxes. The bottom panels plot binned proper motion dispersions as a function of radial distance. We also plot the LOS velocity dispersion and 68\% confidence intervals for each cluster. 

The mean tangential velocity of Praesepe is greater than zero at $\sim 5\sigma$ significance and the binned velocities increase with radial distance. For the Pleiades and M35, the 95\% confidence intervals of the mean tangential velocity and all binned points overlap with zero. M35 shows a signature of rotation about an axis perpendicular to the LOS at 95\% confidence.

The Pleiades show the greatest mean radial discrepancy at a $4.5\sigma$ level. Each binned point displays negative radial motion. The Praesepe mean radial motion is in statistical agreement with zero, but the two outermost bins are $> 3\sigma$ outliers. M35 shows a similar trend of agreement with zero among the inner radial bins, but $2-3\sigma$ discrepancy with the most distant two.

The Pleiades and Praesepe show significant discrepancies between their mean radial and tangential velocity dispersions ($\sim 3\sigma$ and $\sim 6\sigma$ respectively), but for each cluster these values overlap with the 95\% confidence interval for the LOS dispersion. M35 shows the closest agreement among dispersions in all three directions. Among the binned velocity dispersion points, all values overlap with each other and the LOS dispersion at 95\% confidence levels.

\begin{table*}
\begin{center}
\caption{Mean Determinations of Pleiades, Praesepe, and M35 Internal Kinematics.}
\label{tbl:kinematics}
\begin{tabular}{cccccc}
\hline
\hline
 Cluster & Direction & $\theta_{\rm c}$ [$^{\circ}$] & $v_{0}$ [km s$^{-1}$] & $v_{\rm rot}$ [km s$^{-1}$] & $\sigma$ [km s$^{-1}$] \\
\hline
& Tangential & --- & $-0.018 \pm 0.022$ & --- & $0.597 \pm 0.016$ \\
Pleiades & Radial & --- & $-0.106 \pm 0.024$ & --- & $ 0.673 \pm 0.018$ \\
& LOS & $270 \pm 40$ & $5.72 \pm 0.10$ & $0.25 \pm 0.15$ &
$0.73 \pm 0.10$ \\
\hline
& Tangential & --- & $0.132 \pm 0.027$ & --- & $0.611 \pm 0.019$ \\
Praesepe & Radial & --- & $-0.027 \pm 0.036$ & --- & $ 0.811 \pm 0.025$ \\
& LOS & $300 \pm 100$ & $35.000 \pm 0.097$ & 0.093$_{-.065}^{+.10}$ &
$0.626 \pm 0.099$ \\
\hline
& Tangential & --- & $0.010 \pm 0.027$ & --- & $0.808 \pm 0.020$ \\
M35 & Radial & --- & $-0.052 \pm 0.027$ & --- & $ 0.812 \pm 0.020$ \\
& LOS & $200 \pm 30$ & $-8.041 \pm 0.060$ & $0.186 \pm 0.088$ &
$0.848 \pm 0.045$ \\
\hline
\end{tabular}
\end{center}
\tablecomments{$\theta_{\rm c}$ is the position angle of each cluster's LOS rotation axis; $v_{0}$ is the velocity in the specified direction, $v_{\rm rot}$ is the LOS rotation velocity, and $\sigma$ is the velocity dispersion in the specified direction.}
\end{table*}

\begin{figure*}
    \centering
    \includegraphics[scale=0.425]{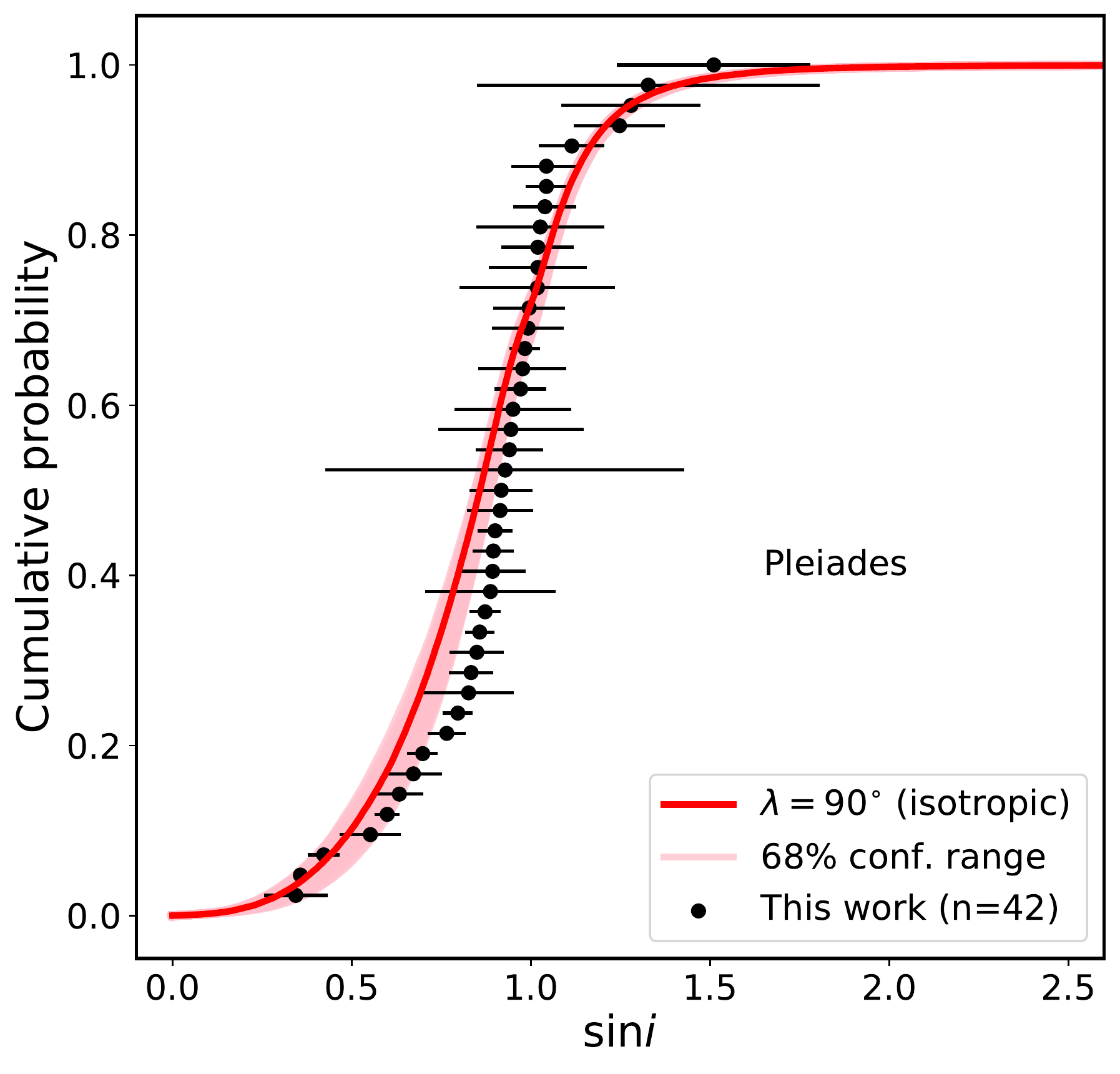}
    \includegraphics[scale=0.425]{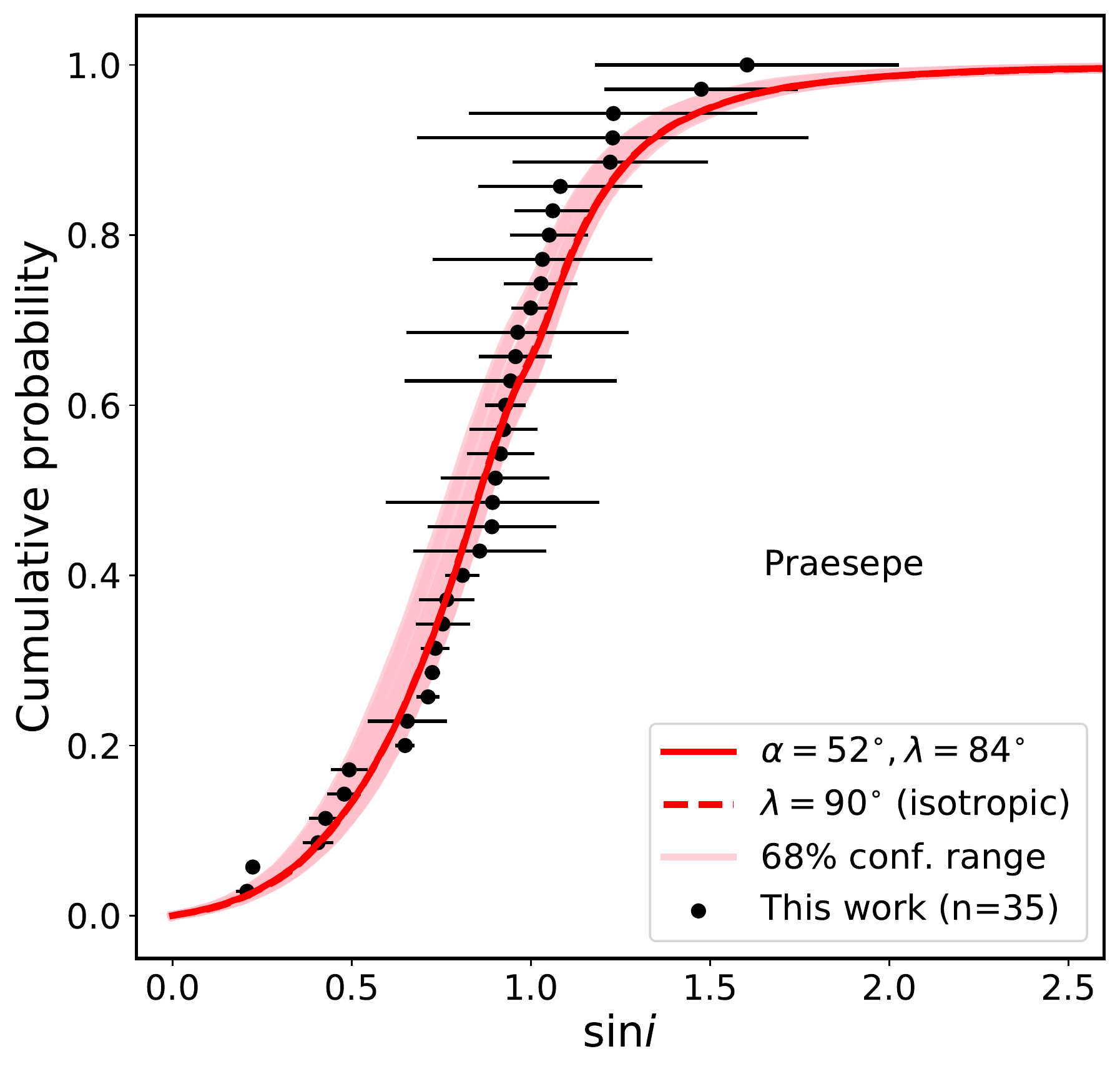}
    \includegraphics[scale=0.425]{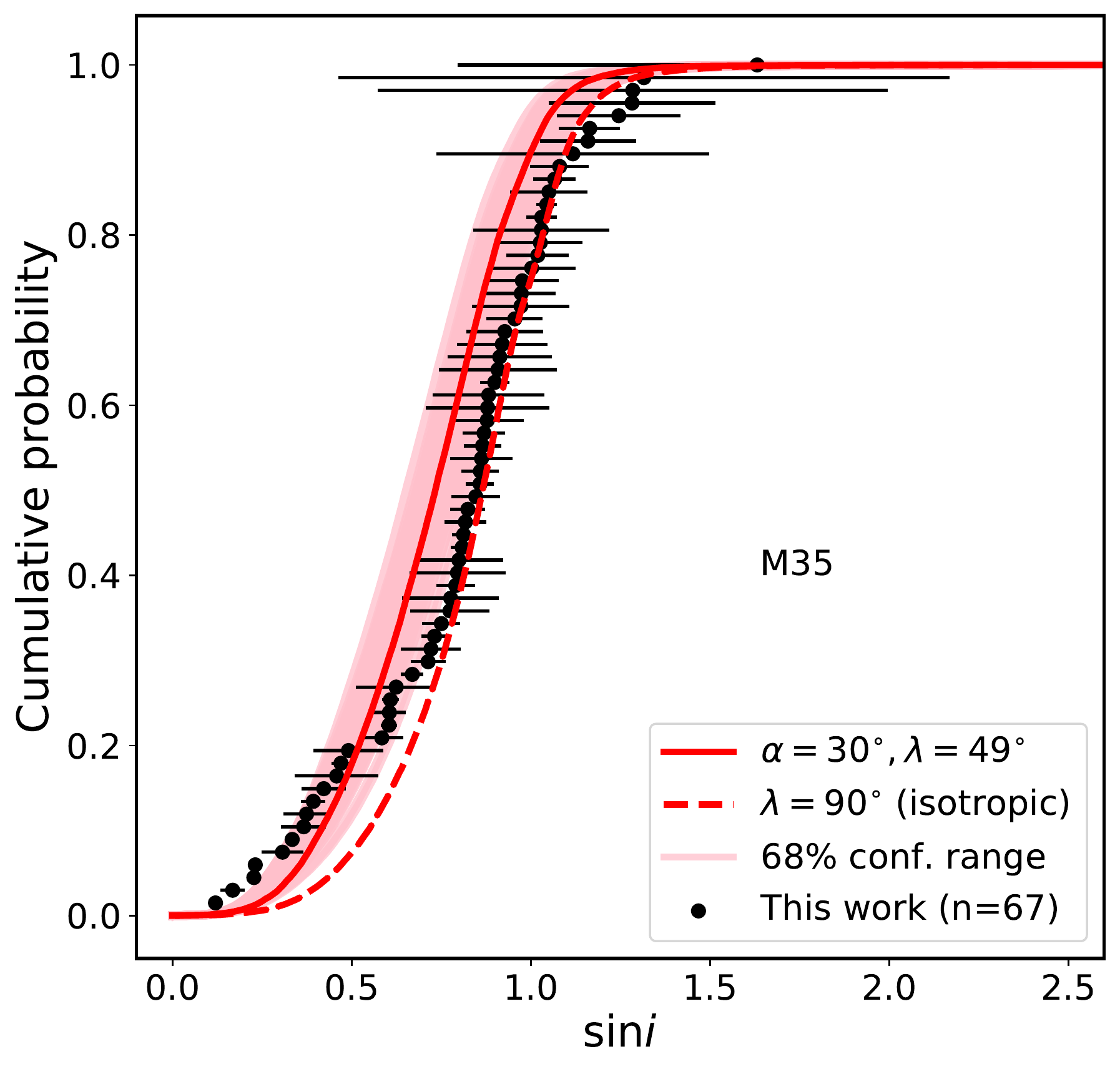}
    \caption{Empirical cumulative distribution functions for  $\sin i$ determinations (black) and the two-parameter model distributions (red) corresponding to the marginalized most probable values of $\alpha$ and $\lambda$. We generated each model distribution using using a Monte Carlo method with $\sim 10^5$ points (see Sec. \ref{subsec:twothreeparam_models}). We include an isotropic model (dashed curve) in panels where it is not already the favored model to facilitate its comparison to the most probable distribution. Pink CDFs represent parameters drawn from the 68\% confidence intervals for $\alpha$ and $\lambda$.}
    \label{fig:ecdf_figs}
\end{figure*}

\begin{figure}
    \centering
    \includegraphics[scale=0.4]{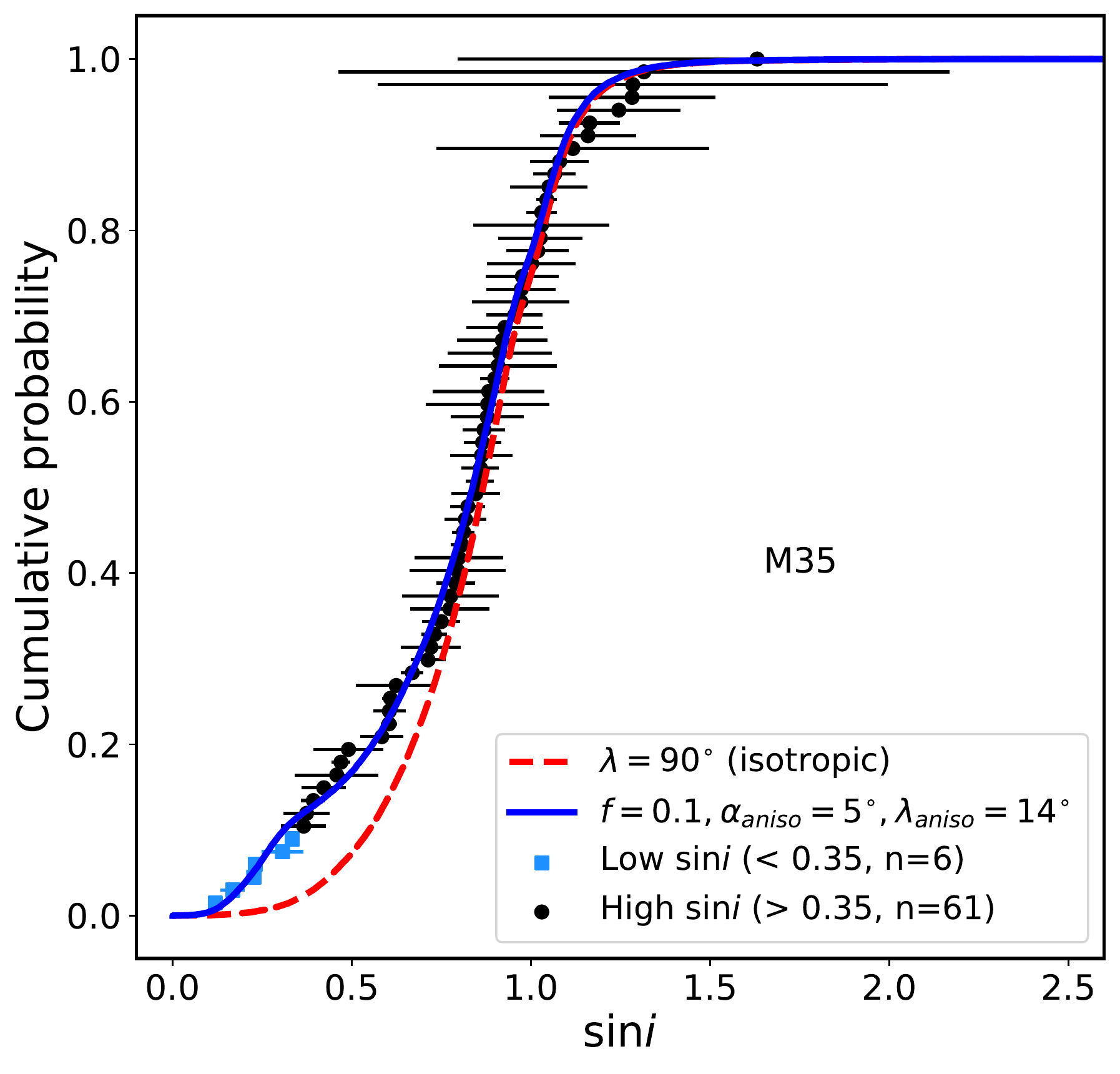}
    \caption{M35 $\sin i$ CDF. We plot the isotropic distribution in red and the best-fitting three-parameter model in blue. The latter model illustrates a scenario in which 10\% of the cluster's stars are tightly aligned nearly along the line of sight, while the rest are isotropically oriented. Blue squares mark the stars having $\sin i < 0.35$, within the inclination range of the anisotropic subset of the three-parameter model.}
    \label{fig:m35_ecdf_twomodels}
\end{figure}

\begin{figure}
    \centering
    \includegraphics[scale=0.45]{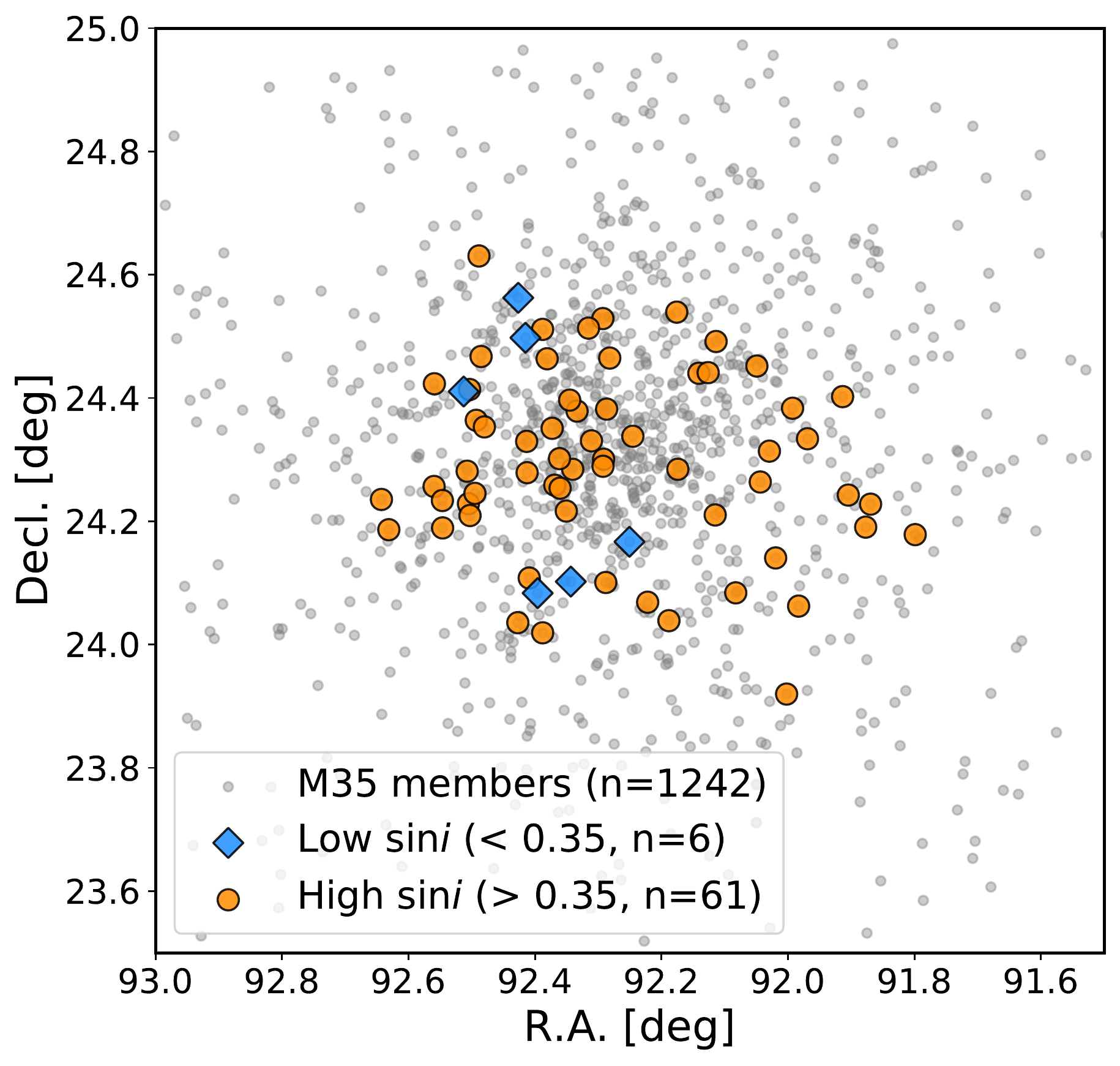}
    \caption{Plot of cluster member positions for M35 (gray). The stars with $\sin i$ determinations are divided along $\sin i = 0.35$, with the 61 stars above this threshold marked by orange circles and 6 below shown as blue diamonds.}
    \label{fig:m35_ra_dec_positions}
\end{figure}

\begin{figure}
    \centering
    \includegraphics[scale=0.45]{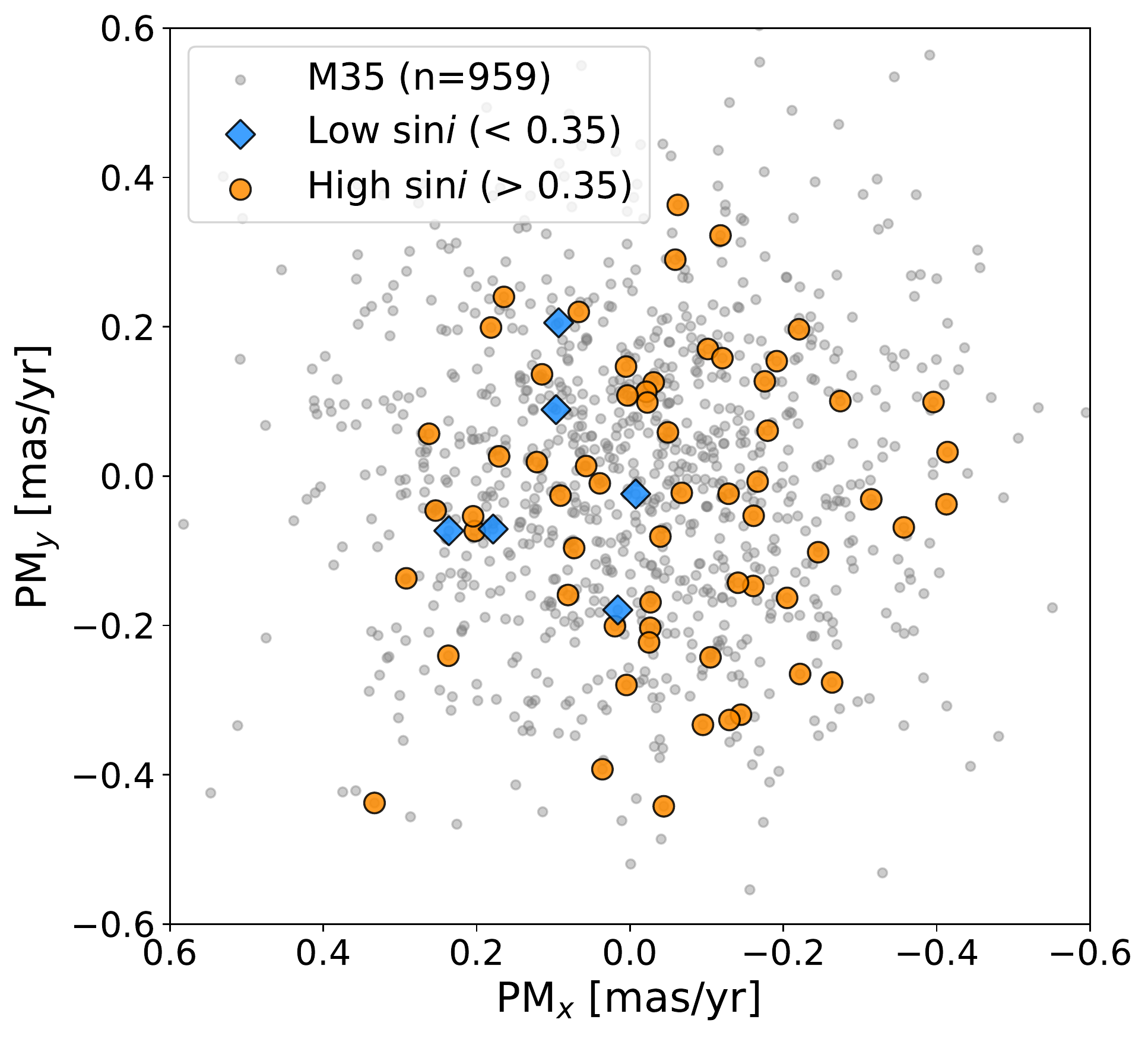}
    \caption{Gaia EDR3 internal proper motions for M35 stars with $\sin i$ determinations, divided into the same groupings as the previous figure. We converted R.A. and decl. into Cartesian coordinates and subtracted expected motion due to the cluster's LOS velocity.}
    \label{fig:m35_corr_pm_x_y}
\end{figure}

\begin{figure*}
    \centering
    \includegraphics[scale=0.45]{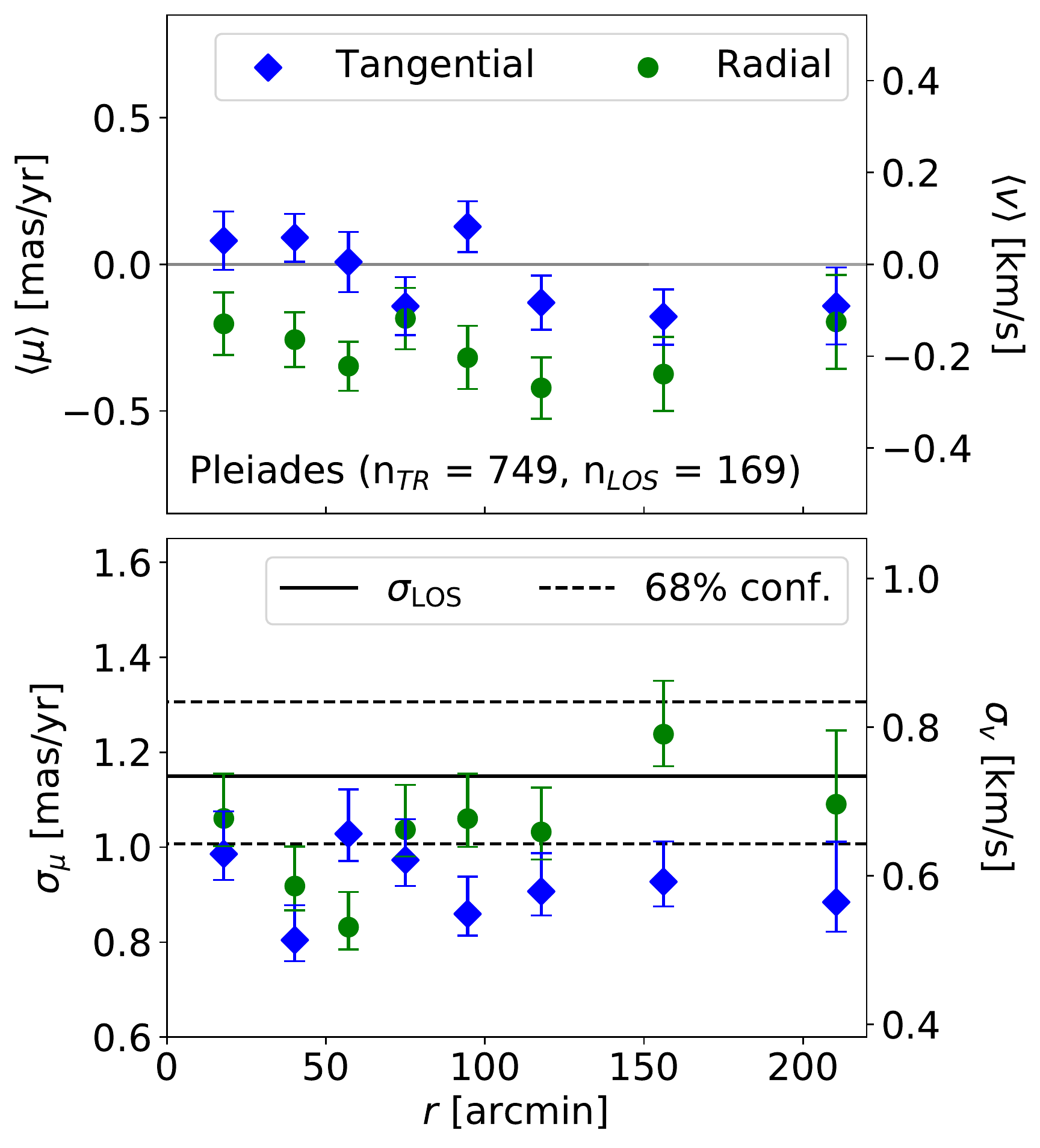}
    \includegraphics[scale=0.45]{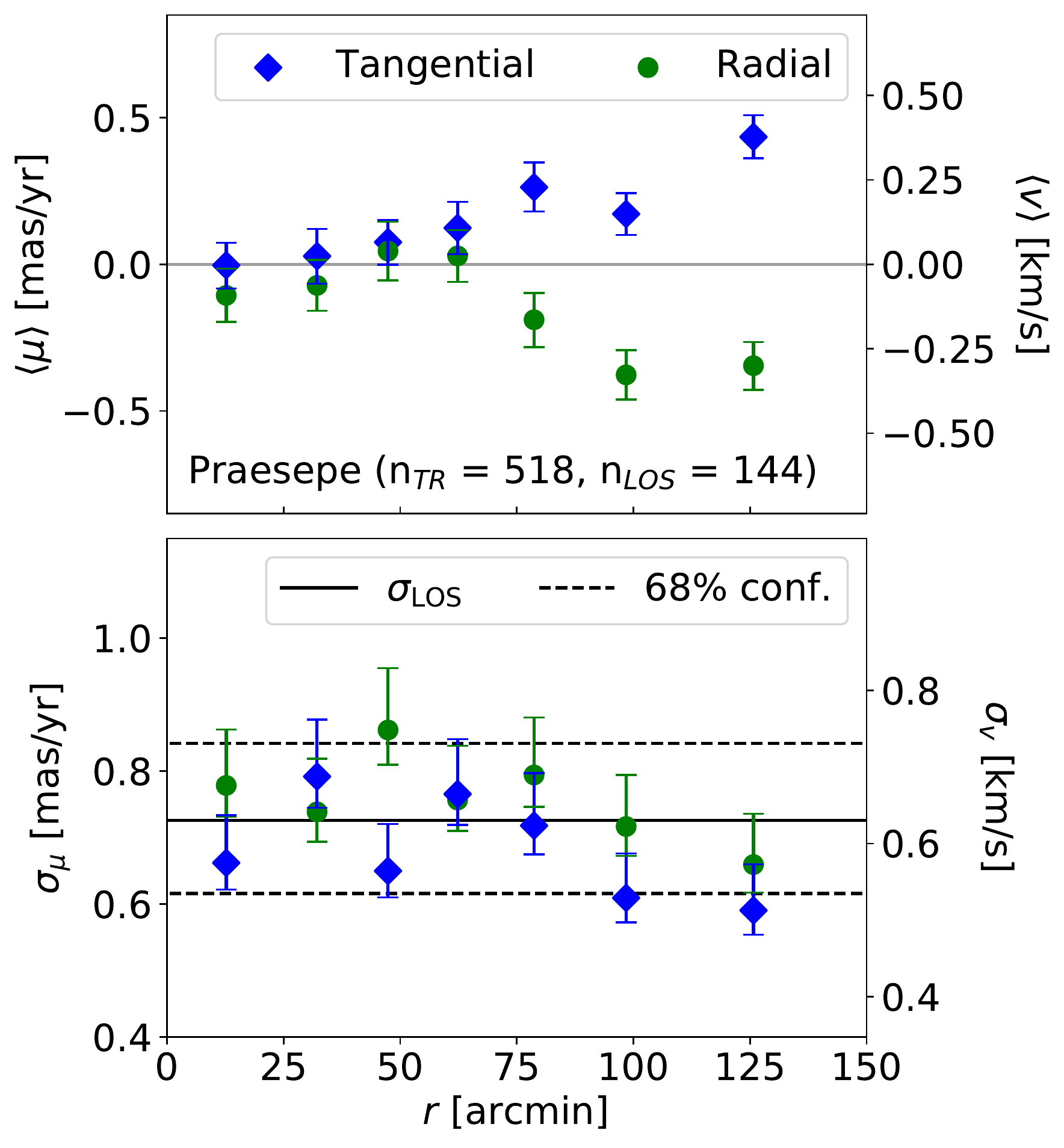}
    \includegraphics[scale=0.45]{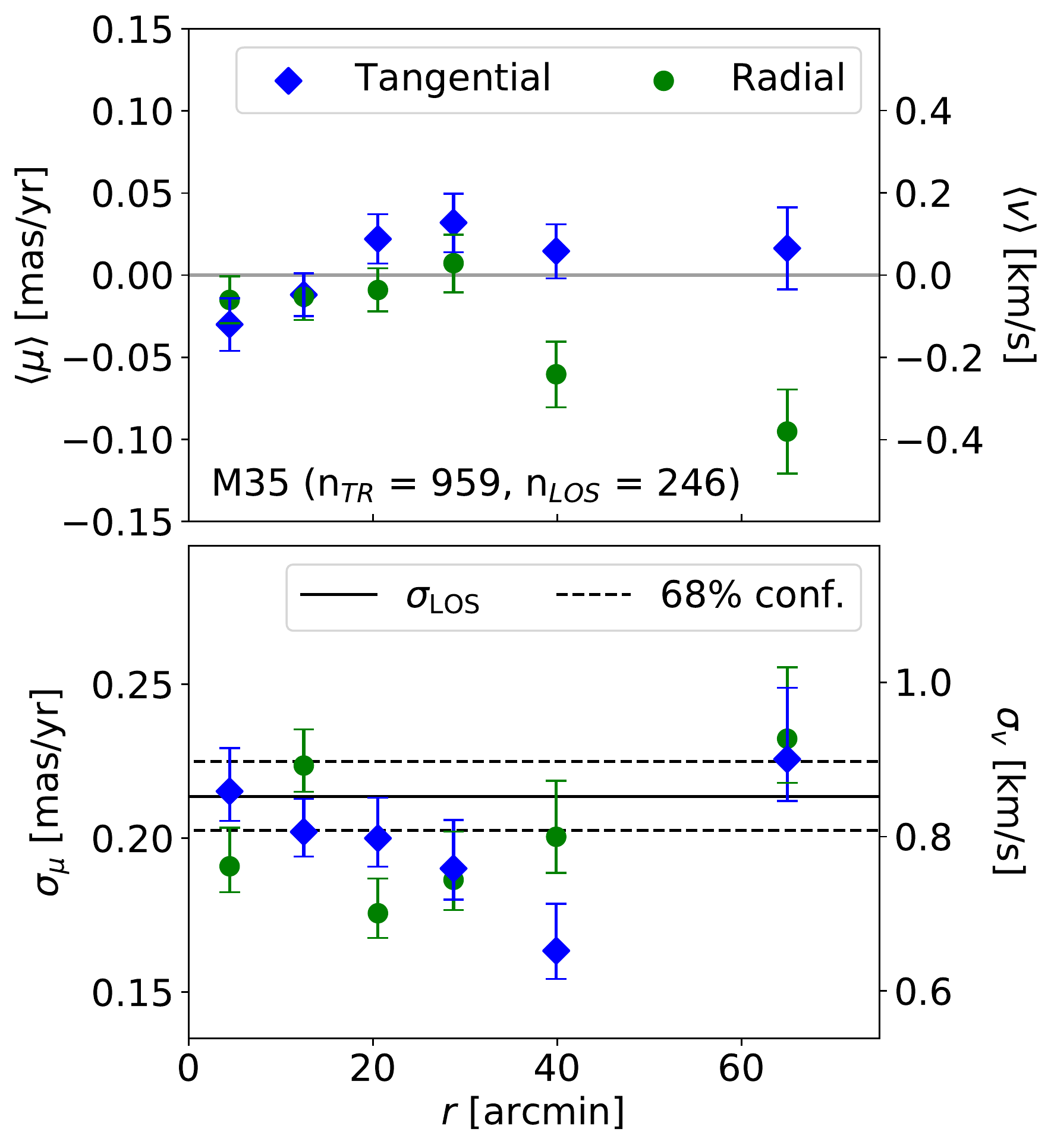}
    \caption{Top panels: binned tangential (blue) and radial (green) proper motions as a function of plane-of-sky distance to cluster center. Expected radial trends from LOS motion have been subtracted. Bottom panels: binned proper motion dispersions as a function of radial distance. We also plot LOS velocity dispersions for each cluster along with 68\% confidence intervals in black. The number of stars analyzed for each cluster's tangential and radial kinematics is displayed as n$_{TR}$, while n$_{LOS}$ indicates the number of LOS velocity measurements.}
    \label{fig:cluster_rot_kinematics}
\end{figure*}

\section{Discussion}
\subsection{Uncertainty distributions and model CDFs}
\label{subsec:error_distribs}
In this section, we discuss how the modeling of uncertainty distributions in the measurements that yield $\sin i$ values (see Eq. \ref{eq:spectrophot}) affect the model CDF shape. In \paperone, we assumed that $v\sin i$, $P$ and $R$ all had uncertainties that were normally distributed. In this work, we approximate the dominant source of error in each cluster's inclination values, $v\sin i$, as a log normal distribution. Forward-modeling the error distributions is important for accurately analyzing an inclination distribution because including the error parameters in the fitting process will introduce degeneracies with the alignment parameters $\alpha$ and $\lambda$, complicating any determination of spin alignment or isotropy \citep[see also Sec. 3 of][]{kovacs2018}.

The choice of uncertainty distribution influences the shape of the resulting inclination models. Simulated $\sin i$ values generated with log-normally distributed errors tend to moderate the curvature of the model CDF compared to inclinations generated from normal error distributions. This facilitates the modeling of $\sin i$ values that suffer from high levels of random error but are nonetheless important to include in the analysis. Being too selective in removing data points based on $\sin i$ (especially those greater than unity) can introduce a systematic bias toward spin alignment.

In each panel of Figure \ref{fig:ecdf_figs}, the model CDFs do not have the same steepness as the black points. However, the error bars on the $\sin i$ values and the 68\% confidence interval CDFs emphasize the low statistical significance of these discrepancies. It is possible that the measurements in this work do not strictly follow a normal or log normal uncertainty distribution, explaining residual disagreement between model and data.

\label{sec:discussion}
\subsection{Pleiades inclinations}
The marginalized model parameters corresponding to the Pleiades inclination distribution are consistent with both isotropy and aligned spins to a moderate degree, similar to NGC 2516 in \paperone. Isotropic or moderately aligned spins in the Pleiades would be consistent with the results reported for GK stars by \citet[][J10]{jackson2010}. We cross matched the J10 sample with Gaia DR2, resulting in 34 stars with 7 overlapping targets between J10 and our sample.\footnote{There is low overlap between the studies because the J10 inclination sample has a mean effective temperature of $\sim 4800 $ K, while our Pleiades targets are hotter with a mean $T_{\rm eff} \sim 5900$ K.} The mean discrepancy between our $\sin i$ values and those of J10 is $\sim 11\%$, comparable to the mean uncertainty of our inclinations. The greatest discrepancy is $\sim 36\%$, which appears to come from a discrepancy in $v\sin i$ near our threshold of 5 km s$^{-1}$.

In addition to the possibility that the uncertainties in Pleiades $v\sin i$ measurements are not log-normally distributed (discussed in Sec. \ref{subsec:error_distribs}), residual discrepancies between the model and data CDFs may be caused by systematic offsets in one or more of the values listed in Equation \ref{eq:spectrophot} that propagate to the final $\sin i$ values. Radius inflation, in which stellar radii are larger than predicted by isochrones, is one possible source of systematic error in $\sin i$. Multiple studies have presented evidence for radius inflation specifically within the Pleiades.

\citet[][]{lanzafame_pleiades_radius_inflation} determined better than 10\% agreement between measured and predicted radii of Pleiades stars with masses between 0.85 and 1.2 \msun, but found deviations at 95\% confidence for stars in the 0.6--0.8 \msun\ range. \citet[][]{jackson2018} found radii inflated by $\sim 15\%$ for Pleiades stars between 0.1 and 0.8 \msun, and their results favor isotropic inclinations while ruling out values of $\lambda < 30^{\circ}$. \citet[][S17]{somers_2017_pleiades_radius_inflation} observed several instances of inflated radii for stars with $T_{\rm eff} < 5700$ K, with minimal offsets for hotter stars (above 6000 K). This latter result highlights the possibility that even some Pleiades stars of G or later spectral type might have inflated radii.

We accessed the radius measurements of S17 and compared them with the MIST isochrone corresponding to our input parameters for Pleiades SED fitting. Depending on the color-to-effective-temperature conversion used in S17, we found that the mean fractional radius difference between the measured radii and the isochrone's prediction is 6--9\% when selecting stars with $T_{\rm eff}$ between 4900 and 5900 K, which approximately corresponds to the cooler half of stars in our $\sin i$ sample for the Pleiades. Between 70\% and 80\% of S17 stars in this effective temperature range have radii greater than our isochrone's prediction. This discrepancy in radii for a large fraction of cooler stars suggests that our radius values determined from SED fitting may be underestimated, thereby overestimating $\sin i$ (see Eq. \ref{eq:spectrophot}). 

The potential systematic effect of radius inflation emphasizes the benefit of accurately quantifying radius systematics in the young clusters of our study. Applying a high-confidence correction to inflated radius measurements could reduce or overcome this systematic error.

\subsection{Praesepe inclinations}
Our analysis of the 35 $\sin i$ values that we determined for Praesepe is most similar to the first, isotropic test that we performed in Sec. \ref{subsec:twothreeparam_models}.
Analyzing a larger sample of 113 stars, \citet[][K18]{kovacs2018} found that the most likely two-parameter model results for the Praesepe inclination distribution were $\alpha = 76^{\circ} \pm 14^{\circ}$ and $\lambda = 47^{\circ} \pm 24^{\circ}$. These parameters are similar (by design) to those of the third and fourth validation runs of our MCMC method in Sec. \ref{subsec:twothreeparam_models}.
As shown by these simulations, a moderately aligned scenario like the one reported by K18 is not statistically distinguishable from isotropy for both 30 and 120 star samples, even when including stars with $v\sin i < 5$ km s$^{-1}$ in the latter.
The minimal dependence of the precision of the parameters on the sample size suggests that the difference in quantity of sini values between this work and K18 is not the source of the discrepant results.

Sources of the smaller number of $\sin i$ values reported in this work compared with K18 include our use of the cluster member list of \citet[][]{cantatgaudin2018}, elimination of the binary main sequence, establishment of a RUWE cutoff to further cull binaries, use of a single \ktwo\ rotation period catalog \citep[][]{rebull2017_praesepe} versus the additional inclusion of ground-based measurements \citep[][]{kovacs2014}, the exclusion of stars with $v\sin i < 5$ km s$^{-1}$ and removal of stars with spurious rotation periods and LOS velocities. In principle, having a larger sample of $\sin i$ values can amplify the effect of systematic errors on the resulting CDF. We made the above selections to reduce the influence of potential systematic error in our inclination distribution. For example, we find evidence of systematics among the equal-mass binary sample in that $\sim 33\%$ of the K18 stars we classify as binaries using the cluster's CMD have $\sin i > 1.75$, compared with $\sim 3\%$ of the remaining 95 stars on the single-star main sequence. K18's removal of stars with $\sin i > 2$ preferentially excludes many of these binaries, achieving a similar result to our explicit exclusion of CMD outliers. 

\begin{figure*}
    \centering
    \includegraphics[scale=0.45]{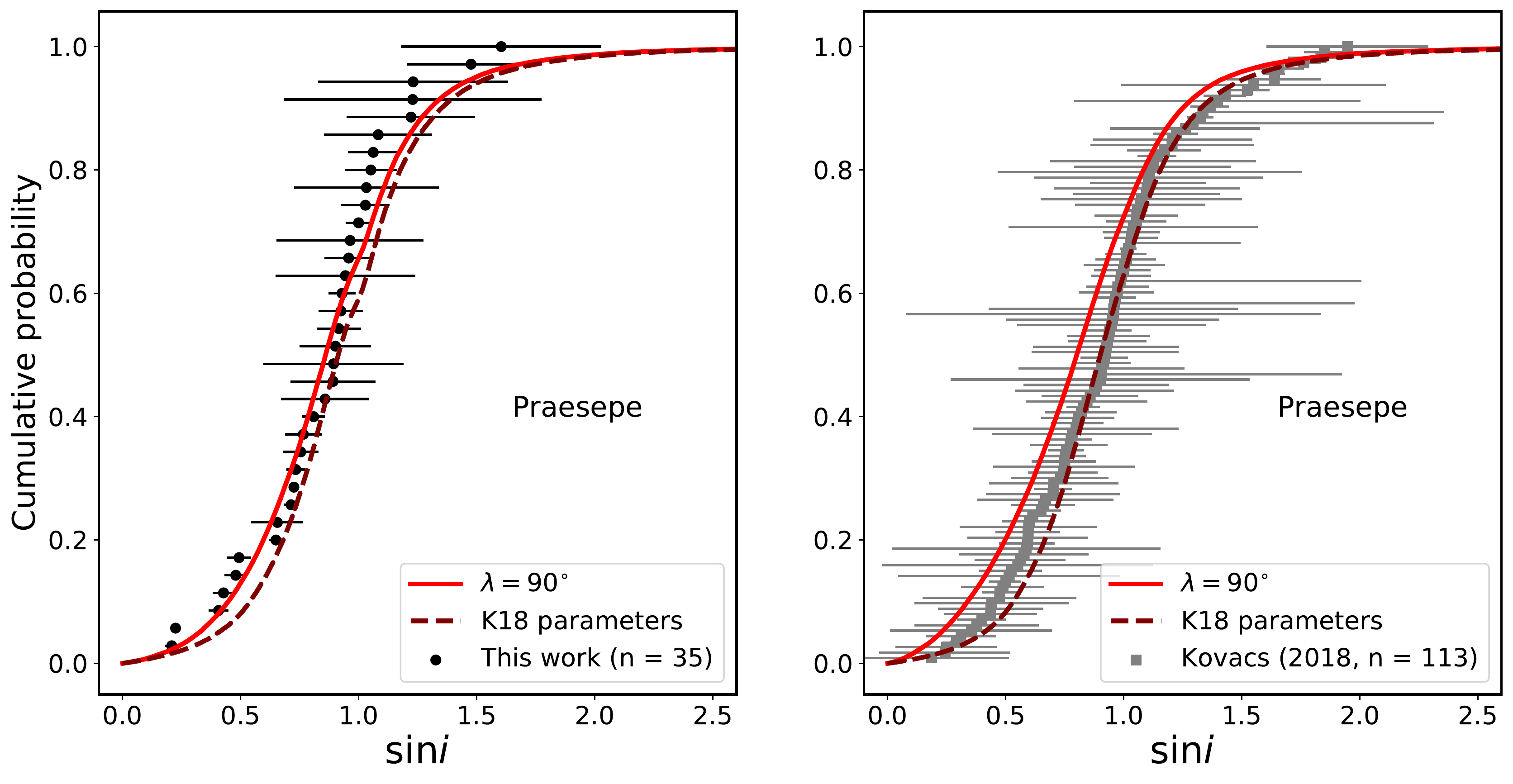}
    \caption{Comparison between this work's $\sin i$ determinations for Praesepe (left) and those of \citet[][right]{kovacs2018}.
    The empirical CDFs are likely to represent the same underlying distribution (K--S $p$-value $\sim 0.9$). The red solid curves in each panel show a model isotropic distribution, while the maroon dashed curves were generated with the parameters reported in K18 ($\alpha = 76^{\circ}$, $\lambda = 47^{\circ}$). At left, the model CDFs account for $v\sin i$ threshold of 5 km s$^{-1}$.}
    \label{fig:praesepe_k18_compare}
\end{figure*}

Figure \ref{fig:praesepe_k18_compare} plots our data in the left panel and those of K18 at right. Both sets of $\sin i$ values have a high probability of being drawn from the same distribution (two-sample K--S test $p$-value $\sim 0.9$). The overall compatibility of the K18 distribution with this work's validates the K18 $\sin i$ determinations. However, our work's determination of the spread half-angle and mean inclination for the cluster does not strongly prefer a moderately aligned scenario over the isotropic case. In each panel of Figure \ref{fig:praesepe_k18_compare}, the red curve represents an isotropic spin-axis distribution, while the maroon dashes represent the K18 values for $\alpha$ and $\lambda$. The similarity of the two distributions underscores the degree of similarity between the two scenarios. We conclude that the possibility of spin alignment as reported by K18 remains, but the distribution is consistent with both isotropy and moderate alignment, much like our results for NGC 2516 in \paperone\ and the Pleiades in this work.

\subsection{M35 inclinations}
\label{subsec:m35_spin_discussion}
The M35 inclination distribution, with the largest sample size of the three clusters, is poorest fit by the two-parameter model. The CDF corresponding to the peak of the marginalized PPDs is offset from the majority of $\sin i$ values, and the empirical distribution of stars with $\sin i \lesssim 0.75$ visibly differs from that of the isotropic model, which otherwise shows closer agreement with points above this threshold than the PPD-peak model (Figure \ref{fig:ecdf_figs}). The spatial positions and proper motions of the six lowest-inclination stars do not display statistically significant incompatibilities with the null hypotheses of uniform position angles and consistency with the proper motions of higher-inclination stars. Nonetheless, we further discuss these stars to narrow down the possible explanations for their grouping.

\subsection{Possible systematics}
\label{subsec:possible_systematics}
It is important to consider potential sources of systematic errors to verify they are not the cause of the excess of low $\sin i$ values. We note that among the $v\sin i$, $R$ and $P$ values for M35, the distribution of measured $v\sin i$ is most discrepant with the predicted distribution for an isotropic scenario based on our simulations from Sec. \ref{subsec:twothreeparam_models} (two-sample K--S $p$-value $\sim 6\times 10^{-5}$). This discrepancy is primarily manifest in a factor of $\sim 2$ overdensity in measured versus predicted $v\sin i$ between 10 and 15 km s$^{-1}$. The simulated $R$ and $P$ distributions, constructed using the cluster's distribution of de-reddened Gaia $BP - RP$ color, are in closer agreement with their empirical counterparts (K--S $p$-values $\sim 0.01$ and $0.25$, respectively).

The WOCS study that sourced the $v\sin i$ measurements for this cluster used a solar spectrum to cross correlate with each M35 spectrum. This could lead to systematic errors in $v\sin i$ given the range of effective temperatures in the WOCS sample ($\sim 4900-7500$ K). The solar macroturbulent velocity ($v_{\textrm{mac},\odot}$) is $ 3.21$ km s$^{-1}$ \citep[][]{doyle2014}. For hotter (6000--6500 K) stars than the Sun, $v_{\rm mac}$ is expected to be $\sim 4-6$ km s$^{-1}$ respectively. When a star's projected rotational velocity is comparable to or less than $v_{\rm mac}$, the two sources of line broadening are difficult to distinguish. Especially for hotter stars with larger $v_{\rm mac}$, this degeneracy may lead to systematically increased WOCS $v\sin i$ values when macroturbulent broadening is interpreted as rotational broadening. However, the 10 km s$^{-1}$ $v\sin i$ threshold for this study reduces this concern.

For the cooler and younger than solar stars in the WOCS sample ($\sim 4900 - 5700$) K, macroturbulent broadening is slightly smaller than in the Sun, but the level of magnetic activity is higher and may broaden spectral lines to a greater extent than for the Sun. For example, a comparison between Figure 3 of \citet{schrijver2012_sun_flare_frequency} and Figure 4 of \citet{ilin2019_km_flares_m45} shows that a sample of 737 mostly K--M dwarfs in the Pleiades had flare frequencies nearly two orders of magnitude higher than that of the Sun. Without accounting for the increased activity broadening in young, cool stars, broadening relative to a solar spectrum may be interpreted as rotational in origin during spectral cross correlation, potentially leading to overestimating $v\sin i$ for the cooler M35 stars. This effect may explain why some stars appear above the contour of Figure \ref{fig:per_teff_plots}, where by design we anticipate no stars should appear because of the 10 km s$^{-1}$ threshold applied to $v\sin i$ for M35.

Due to the similar young ages of M35 and the Pleiades, it is possible that radius inflation exists among the cooler stars of both clusters. Since we are currently using radius values not corrected for possible inflation, we are determining an upper limit for $\sin i$ rather than a lower limit (see Eq. \ref{eq:spectrophot}). Thus, in the case of the low $\sin i$ values of interest in M35, it is unlikely that radius inflation is the source of the discrepancy with isotropy.

The effects described above would manifest in increased $\sin i$ values above their actual inclination, but they do not explain a distribution that has an overdensity of lower inclinations. Aliased rotation periods can cause integer discrepancies in the resulting $\sin i$ values, yielding significant systematics in both directions. However, we have already removed periods that are highly discrepant from empirical color--period isochrones, and we have visually inspected the light curves of each star to verify that the periods appear to be fundamentals, rather than harmonic.

While we suspect that $v\sin i$ is systematically overestimated for the cooler stars of the sample, an opposing systematic would have to be acting to decrease $v\sin i$ measurements for other stars, and it would have to be happening across the CMD - the six stars of the low-inclination subset range in $T_{\rm eff}$ from $\sim$ 5000 to 7000 K, spanning both the slow- and fast--rotating sequences of the cluster. It remains possible that an unexplained systematic error in $v\sin i$ affecting measurements near the 10 km s$^{-1}$ threshold is responsible for the excess of low-inclination values.

\subsection{Differential rotation}
\label{subsec:diffrot}
Differential rotation can lead to systematically longer period measurements when spots at nonzero latitudes are moving slower than their star's equatorial rotation velocity. The measurement of $v\sin i$ can also be underestimated due to this effect since a differentially rotating star will have less of its surface contributing to extreme line-of-sight velocities compared to a uniform rotator.

Sec. 3.2.3 of \citep[][]{hirano2014} considered the combined effect that these biases can have on inclination determinations. The authors quantified differential rotation using the equation
\begin{equation}
    \Omega(l) = \Omega_{\rm{eq}}(1 - \alpha \sin^2{l}),
    \label{eq:diffrot}
\end{equation}
where $\Omega(l)$ is the angular velocity at latitude $l$, $\Omega_{\rm{eq}}$ is the equatorial angular velocity, and $\alpha$ is the ratio of polar to equatorial angular velocities. By combining Eq. \ref{eq:diffrot} with an empirical relation between $\alpha$, $\Omega_{\rm{eq}}$ and a young star's effective temperature \citep[Figure 1 of][]{colliercameron2007_diffrot}, it is possible to estimate the amount by which rotation periods are measured systematically longer. We calculate that for the stars of the Pleiades, Praesepe, and M35, $\alpha$ is approximately 0.1-0.2. Combined with the assumption that spots are typically situated at Sun-like latitudes of $|l| \sim 20^{\circ} \pm 20^{\circ}$, we find that measured rotation periods could be overestimated by 1\%--2\% on average, and 4\%--8\% at $l=40^{\circ}$.

By simulating changes to a spectral line profile, \citet[][]{hirano2014} also approximated the effect of differential rotation on projected rotation velocity measurements. Their modeling estimated that the fraction $f$ by which $v\sin i$ is underestimated is $f \sim 1-\alpha/2$. For $\alpha$ between 0.1 and 0.2, this yields a 5--10\% fractional underestimate of $v\sin i$.

In Eq. \ref{eq:spectrophot}, $\sin i$ is proportional to both $P$ and $v\sin i$. Thus, an overestimated period and an underestimated rotation velocity will act in opposing directions to change the determined inclination. Using the quantitative estimates we have described above, we predict the underestimation of $\sin i$ by a factor between 1\% and 8\% due to differential rotation. Since the typical random $\sin i$ uncertainties for each cluster in this work dominate the potential systematic error in quadrature, differential rotation cannot significantly affect the results of our analysis. This effect is also not a satisfactory resolution to the excess of low-inclination stars in M35.

\subsection{Physical interpretations of M35 grouping}
\label{subsec:m35_physical_insights}
A possible physical interpretation of the low $\sin i$ grouping is that the stars share similar inclinations because they inherited angular momentum from a clump that formed within the larger progenitor giant molecular cloud. \citet[][and references therein]{degris_star_formation_review} describe the observations and theory that suggest a prevalence of clumps and other substructures in molecular clouds that form clusters. Simulations such as those by \citet[][]{federrath2013} find that within a molecular cloud, stars form mostly in dense filaments.

If such a filament or clump explains the low $\sin i$ grouping in the 150 Myr old M35, it would be necessary to explain why the phases of the stars' orbits did not get randomized after roughly 8 crossing times ($\sim 18$ Myr for M35). If the stars are gravitationally bound and orbit the cluster center with the same period, they could avoid the aforementioned dispersal of their spatial grouping. However, while the relaxation time for M35, at $t_{\rm relax} \sim 570$ Myr, is greater than its age, cluster substructure is expected to be lost on a timescale of a few Myr \citep[e.g.][]{schmeja2008}. This prediction contradicts the interpretation of the low-$\sin i$ stars as a grouping related from their formation. It is also possible that these stars more recently merged with M35, having formed nearby but separately from the rest of the cluster.

Confidence in these stars' physical association could be increased by determining the positions of each star along the LOS relative to cluster center. For example, if all are located on the same side of the cluster, it would be more likely that they are physically associated. While EDR3 parallax measurements cannot provide this insight, a K--S test suggests that the distribution of LOS velocities of low $\sin i$ stars is not significantly different than that of high $\sin i$ stars. Contrasted with the significant groupings of position angles and proper motions for low $\sin i$ stars, the data do not offer a conclusive answer on their physical association.

If the low-inclination stars are indeed physically associated, their alignment would be consistent with the results of multiple numerical simulations.  \citet[][]{corsaro2017} simulated a molecular cloud with radius 0.084 pc and mass $10^3$ \msun\ forming a 140 \msun\ proto-cluster. The simulations predict that if $\sim 50\%$ of a molecular cloud's kinetic energy is in overall rotation, then the transfer of global angular momentum can result in aligned spins for stars $> 0.7$ \msun. Based on their effective temperatures and Table 5 of \citet[][]{peacut2013_spt_mass_G}, all of the low $\sin i$ stars in M35 have masses greater than this threshold. Below 0.7 \msun, not enough material is accreted to overcome the randomizing effect of turbulence, which would contribute most of the remaining $50\%$ of the cloud's kinetic energy.\footnote{This result cautions against the extrapolation of outcomes from, e.g., \citet[][]{jackson2018} and \citet[][]{jackson2019}, which suggested isotropic inclinations among low-mass stars in the Pleiades and Praesepe, to the more massive stars of these clusters.} This energy balance is similar to what was found to produce alignment on a larger spatial scale (resolution 0.1 pc) by the simulations of \citet[][]{reyraposo2018}.
These authors suggest that the mixed results from observational studies of spin alignment could all be consistent if the alignment is stronger in higher-mass clusters. While M35 is the most massive cluster of our sample, there is not yet a large enough collection of cluster inclination distributions to firmly establish a trend.

Both of the above simulations neglected magnetic fields, which are known to have an important role in shaping molecular clouds and moderating the star formation rate \citep[][and references therein]{hennebelle2019_magnetic_fields_clouds_review}. One set of magnetized cloud collapse simulations showed that filamentary structures were enhanced by a strong magnetic field, and that these structures were roughly aligned with the large-scale magnetic field lines \citep[][]{price2008}. Stronger magnetic fields in these simulations also resulted in anisotropic turbulent motions, suggesting that even the mechanism expected to randomize stellar spins in clusters could have a reduced influence in some cases.

\subsection{Cluster kinematics}
Each cluster's radial proper motions include binned values that significantly disagree with zero after accounting for the cluster's motion along the LOS (Figure \ref{fig:cluster_rot_kinematics}). The greatest discrepancies with zero occur in the negative (radially inward) direction. The Pleiades show negative discrepancies at all radial distances from cluster center. For Praesepe and M35, the inward discrepancies are pronounced at larger radial distances. Based on the known evolutionary stages of clusters, one might instead expect to observe net motion in a radially outward direction within a cluster's halo, owing to the plentiful low-mass stars that rise (or even escape) as a result of dynamical interactions with more massive stars (mass segregation). It is possible that these discrepancies manifest in each cluster's mean proper motions due to the combined effects of escaped stars and the selection of members with a Gaia $G$ magnitude $< 18$ \citep[][]{cantatgaudin2018}.

The processes of stellar ejection (due to a single close encounter) and, more commonly, evaporation (the sum of numerous weak encounters) cause a cluster to lose members over time \citep[][]{gerhard2000_cluster_dynamics_review}. The numerical simulations of \citet[][]{moyano2013_cluster_escapevelocity} modeled the dynamical evolution of clusters with initial masses ranging from $\sim$ 6000 to 9000 \msun. At ages between 50 and 1000 Myr, these simulations predicted a single-star escape velocity of $\sim 2$ km s$^{-1}$. Defining an escaped star as unbound and surpassing a 100 pc radial distance to cluster center, the simulations found that $\sim$ 20\% of each model cluster's stars escaped within 100 Myr, increasing to $\sim$ 40\% by 600 Myr. In our work, we did not analyze stars at such large distances from their cluster's center.

For simulated clusters aged $\lesssim$ 1 Gyr, most of the escaped stars had mass $< 0.7$\msun. Given the ages of the Pleiades, Praesepe, and M35, between 20\% and 40\% of each cluster's initial stellar population may have escaped, leaving behind stars more massive than 0.7 \msun\ being sent inward toward cluster center due to dynamical interactions with escaping stars. By weighting the outlying radial velocities in Figure \ref{fig:cluster_rot_kinematics} with a mean $\sim 2$ km s$^{-1}$ escape velocity for the predicted fraction of missing stars, we obtain signatures of net expansion, rather than contraction, in all radial bins.

The membership magnitude threshold requiring $G < 18$ may also bias the radial kinematics. We compute the absolute Gaia magnitude $M_{G}$ of the faintest possible star to be included in each cluster's list of members and use the results of \citet[][]{cifuentes_2020_gmag_mapping} to map $M_{G}$ to an approximate spectral type. We find that for the Pleiades and Praesepe, the $G$-magnitude threshold results in the exclusion of main-sequence stars later than M4.5 and M4, respectively. For the more distant M35 cluster, the spectral type cutoff is K8. According to Table 5 of \citet{peacut2013_spt_mass_G}, these spectral types correspond to masses of $\sim$ 0.18, 0.23, and 0.62 \msun\ respectively.

Comparing our work's mass thresholds to the mass functions of \citep[][Pleiades and Praesepe]{pleiades_praesepe_mass_functions} and \citep[][M35]{m35_mass_function} suggests that $\sim 30\%$ of cluster members in the Pleiades and Praesepe (and $\sim 60\%$ in M35), all having masses $\sim 0.6$ \msun\ or less, are excluded by the $G$-magnitude cutoff. Assuming that stellar escape is ongoing in these clusters, we exclude these low-mass stars that make up the escaping population. The remaining sample may be biased toward bound stars having inward-directed radial motion, reflecting the results of our analysis for each cluster. This potential bias cautions against interpreting the results of Figure \ref{fig:cluster_rot_kinematics} as representative of the entire cluster.

While spectral type might be expected to predict radial motion in a cluster, there are no expectations for its correlation with net tangential motion, allowing for a less biased determination of cluster rotation. We observe a significant signal of rotation in the plane of the sky only among the stars of Praesepe. The positive mean tangential proper motion corresponds to clockwise rotation.\footnote{Since this paper's coordinate system defines the eastward direction as positive, the rotation direction inferred from the sign of tangential motion is opposite the convention of \citet[][]{kamann2019}.} Even though the relaxation time of Praesepe, at $\sim 140$ Myr, is much less than its age, the rotation signal is significantly nonzero at large distances from cluster center. It may be possible that relaxation has dissipated angular momentum closer to the center, but not in the outer reaches of the cluster. Another Gaia-based study, \citet{loktin_praesepe_rotation}, also found a plane-of-sky signature of rotation in Praesepe at large distances from cluster center with a speed of $\sim 0.4$ km s$^{-1}$. This value is quite similar to the tangential speed of the most radially distant bin in our analysis, and both papers report the same direction of rotation. The other two clusters do not show significant signs of rotation in their proper motions. 

We also see a less significant rotation signal about an axis perpendicular to the LOS in M35. The younger M35 (age $< t_{\rm relax}$) is less likely to have experienced dissipation of its primordial angular momentum. Since we do not detect significant rotation in the plane of the sky for M35, the inclination of this cluster's rotation axis is likely closer to $90^{\circ}$ (edge-on) than $0^{\circ}$ according to our convention. We make a simple estimate of the rotation axis orientation using the $\arctan$ of the ratio of LOS to plane-of-sky rotation, finding this angle to be $\sim 87^{\circ}$. Therefore, M35's rotation axis is not likely to be consistent with the orientation of stars in the low $\sin i$ grouping, which has a nearly pole-on mean inclination of 5$^{\circ}$. This comparison suggests that if the grouping of stars is indeed a physical substructure in the cluster, its angular momentum is not correlated with that of the cluster as a whole.

According to \citet[][]{cantatgaudin2021_pm_correction}, color-dependent proper motion systematics of $\sim 10\ \mu$as yr$^{-1}$ remain after applying their correction to the Gaia EDR3 data. This level of residual systematics cannot explain the statistically significant deviations from zero that we report. We note that for all clusters, the inferred rotation velocities are at least four times smaller than the velocity dispersion in each dimension. Thus, each cluster is dominated by dispersion rather than ordered rotation. 

\section{Conclusions and Future Work} 
\label{sec:conclusion}
We extended the study of stellar spin-axis orientations that we began with NGC 2516 (\paperone) to three additional open clusters: Pleiades, Praesepe, and M35. Studying many clusters in a systematic way can reveal interesting trends and outliers that offer insight beyond individual clusters. 

In the Pleiades and Praesepe, we found inclination distributions consistent with both isotropy and moderate alignment. Our result for the Pleiades agrees with a previous study by \citet[][]{jackson2010}. We did not find evidence of overall cluster rotation in the Pleiades.

We compared our Praesepe results to those of \citet[][]{kovacs2018}, who reported moderate spin alignment in the cluster. We found that both studies' inclination distributions closely overlap, but our analysis of the distribution does not prefer alignment over isotropy.
While our analysis does not rule out moderately aligned scenarios, its results are consistent with simulations that modeled isotropic spins. We also observed a significant signature of this cluster's overall rotation in the plane of the sky.

The addition of more inclination distributions consistent with both isotropy and moderate alignment to our sample of clusters begins to provide a path toward breaking the degeneracy associated with each one. Assuming that the direction and degree of spin alignment is not correlated across multiple clusters, it is more probable that a collection of inclination distributions with degenerate parameters represents isotropically oriented spins instead of similar moderate alignment and mean inclinations for all clusters in the sample. Our planned study of $\sim 10$ more clusters may help us to infer isotropic spins for a larger collection of inclinations which, on their own, do not offer such conclusive results.

Our largest sample of inclination determinations in M35 was overall best fit by a moderately aligned model, while the isotropic model shows closer agreement with more points in the CDF. However, within the distribution of M35 $\sin i$ values, we identified an overdensity of stars preferentially inclined close to the LOS, six of which showed visual (but statistically insignificant) clustering of proper motions and position angles. We did not identify a systematic effect that could result in this grouping of inclinations, but among the measurements in Eq. \ref{eq:spectrophot} used to determine $\sin i$, the $v\sin i$ distribution was most discrepant with isotropic predictions. It is possible than an unexplained systematic error in $v\sin i$ values near the measurement threshold is the source of this discrepancy. We also explored the physical interpretation that the grouping is composed of associated stars that inherited common angular momentum at the time of their formation in a clump with a roughly equal balance of rotational and turbulent kinetic energy, aided by organized magnetic fields. At an age of $\sim$ 150 Myr, however, M35 is too old to be expected to display this kind of grouping.

We offer the following ways to further improve future studies: the sample sizes of clusters and member stars could be increased with multi-object spectroscopic observations to determine $v\sin i$ for additional cluster members. The upcoming PLATO mission, with $\sim 30\%$ better angular resolution and $\sim 30$x longer dwell time compared with TESS, will yield precise photometry of stars in more distant and older clusters. While the spectro-photometric method is not favorable for older clusters, PLATO will facilitate new asteroseismic studies of stellar spin through its long-baseline observations, especially for evolved stars.

Models of radius inflation in young stars could improve the accuracy of subsequent $\sin i$ determinations or assist in culling particular stars from the sample. Additional numerical simulations of molecular cloud collapse and cluster evolution could further address the possibility of aligned spins, including whether alignment might manifest and persist in a subgroup of cluster stars. Finally, asteroseismic determinations of stellar inclinations can be less susceptible to the systematic errors inherent to the spectro-photometric method, but reliable measurements require continuous, multi-year photometry at high precision. We will continue our systematic, spectro-photometric open cluster spin study in upcoming work, using publicly available spectro-photometric data to determine the inclination distributions of stars in $\sim 10$ more clusters and synthesizing all of our results.

\acknowledgments

We thank Carlos Allende Prieto, Annalisa Calamida, Eric Feigelson, Geza Kovacs, Luisa Rebull, Henrique Reggiani, David Sing, David Soderblom, Jamie Tayar, and Nadia Zakamska for helpful advice and correspondence. We thank the anonymous referee for helpful comments which strengthened this paper. Some of the data presented in this paper were obtained from the Mikulski Archive for Space Telescopes (MAST) at the Space Telescope Science Institute. The specific observations analyzed can be accessed via\dataset[10.17909/T93W28]{https://doi.org/10.17909/T93W28}.

This paper includes data collected by the Kepler mission and obtained from the MAST data archive at the Space Telescope Science Institute (STScI). Funding for the Kepler mission is provided by the NASA Science Mission Directorate. STScI is operated by the Association of Universities for Research in Astronomy, Inc., under NASA contract NAS 5?26555. We acknowledge support from the STScI Director's Research Funds under grant No.\ D0101.90264.

This work has made use of data from the European Space Agency (ESA) mission
{Gaia} (\url{https://www.cosmos.esa.int/gaia}), processed by the {Gaia}
Data Processing and Analysis Consortium (DPAC,
\url{https://www.cosmos.esa.int/web/gaia/dpac/consortium}). Funding for the DPAC
has been provided by national institutions, in particular the institutions
participating in the {Gaia} Multilateral Agreement.

This publication makes use of data products from the Two Micron All Sky Survey, which is a joint project of the University of Massachusetts and the Infrared Processing and Analysis Center/California Institute of Technology, funded by the National Aeronautics and Space Administration and the National Science Foundation. This publication makes use of data products from the Wide-field Infrared Survey Explorer, which is a joint project of the University of California, Los Angeles, and the Jet Propulsion Laboratory/California Institute of Technology, and NEOWISE, which is a project of the Jet Propulsion Laboratory/California Institute of Technology. WISE and NEOWISE are funded by NASA. This paper makes use of data from the AAVSO Photometric All Sky Survey, whose funding has been provided by the Robert Martin Ayers Sciences Fund and from the NSF (AST-1412587).

This research has made use of the SIMBAD database, operated at CDS, Strasbourg, France. This research has made use of the VizieR catalog access tool, CDS, Strasbourg, France. The original description of the VizieR service was published in A\&AS 143, 23. This work made use of NASA's Astrophysics Data System Bibliographic Services. This research has made use of the SVO Filter Profile Service (\url{http://svo2.cab.inta-csic.es/theory/fps/}) supported from the Spanish MINECO through grant AYA2017-84089.

\facilities{\ktwo\ \citep{borucki2010,k2}, Gaia \citep{gaia2016,gaia2018,gaia_edr3_2020}, CTIO:2MASS \citep{skrutskie2006}, WISE \citep{wise_wright2010, catwise_2020}, {Hipparcos}:Tycho-2 \citep{perryman1997,hoeg2000}, AAVSO:APASS \citep{apass_dr10},
GALEX \citep[][]{galex}, LAMOST \citep[][]{cui2012}}

\software{\texttt{astropy} \citep{astropy2018},
\texttt{astroquery} \citep{ginsburg2019},
\texttt{corner} \citep{corner},
\texttt{eleanor} \citep{feinstein2019},
        \texttt{emcee} \citep{foremanmackey2013},
        \texttt{iPython} \citep{ipython},
        \texttt{isochrones} \citep{morton2015},
        \texttt{lightkurve} \citep{lightkurve2018},
        \texttt{matplotlib} \citep{matplotlib},
        \texttt{multinest} \citep{feroz2009_multinest},
        \texttt{ndtest},
        \texttt{numpy} \citep{numpy1,numpy2},
        \texttt{pandas} \citep{pandas},
        \texttt{pymultinest} \citep{buchner2014_pymultinest},
        \texttt{scipy} \citep{scipy},
        \texttt{statsmodels} \citep{statsmodels},
        \texttt{TESScut} \citep{tesscut},
        \texttt{TOPCAT} \citep{taylor2005_topcat},
        \texttt{uncertainties} \citep{uncertainties},
        \texttt{zero-point}
        \citep{lindegren2020_gaia_edr3_parallax_bias}
          }

\bibliography{sample63}{}
\bibliographystyle{aasjournal}

\end{document}